\documentclass[prd,amsmath,superscriptaddress,twocolumn]{revtex4}

\usepackage{hyperref}
\usepackage{ifthen}
\usepackage{amsmath}
\usepackage{amssymb}
\usepackage{graphics}
\usepackage{color,bm}

\definecolor{darkgreen}{cmyk}{0.85,0.2,1.00,0.2}
\definecolor{purple}{cmyk}{0.5,1.0,0,0}

\newcommand{\bq}{\begin{equation}}
\newcommand{\eq}{\end{equation}}
\newcommand{\bqa}{\begin{eqnarray}}
\newcommand{\eqa}{\end{eqnarray}}

\newcommand{\Sm}{S_{\rm m}}
\newcommand{\Om}{\Omega_{\rm m}}
\newcommand{\scal}{\varphi}
\newcommand{\dscal}{\delta\varphi}
\newcommand{\vm}{V_{\rm m}}
\newcommand{\dm}{\Delta_{\rm m}}
\newcommand{\phim}{\Phi_-}
\newcommand{\phip}{\Phi_+}
\newcommand{\rmd}{{\rm d}}
\newcommand{\rhm}{\rho_{\rm m}}
\newcommand{\gsh}{g_{\rm SH}}
\newcommand{\gqs}{g_{\rm QS}}
\newcommand{\cG}{c_{\Gamma}}

\begin{document}

\title{Relativistic effects in galaxy clustering in a parametrized post-Friedmann universe}

\author{Lucas~Lombriser}
\affiliation{Institute of Cosmology \& Gravitation, University of Portsmouth, Portsmouth, PO1 3FX, UK}
\author{Jaiyul~Yoo}
\affiliation{Institute for Theoretical Physics, University of Zurich, Winterthurerstrasse 190, CH-8057 Z\"urich, Switzerland}
\affiliation{Lawrence Berkeley National Laboratory, University of California, Berkeley, CA 94720, USA}
\author{Kazuya~Koyama}
\affiliation{Institute of Cosmology \& Gravitation, University of Portsmouth, Portsmouth, PO1 3FX, UK}

\date{\today}

\begin{abstract}

We explore the signatures of quintessence and modified gravity theories in the relativistic description of galaxy clustering within a parametrized post-Friedmann framework.
For this purpose, we develop a calibration method to consistently account for horizon-scale effects in the linear parametrized Post-Friedmann perturbations of minimally and nonminimally coupled scalar-tensor theories and test it against the full model-specific fluctuations.
We further study the relativistic effects in galaxy clustering for the normal and self-accelerating branches of the Dvali-Gabadadze-Porrati braneworld model as well as for phenomenological modifications of gravity.
We quantify the impact of modified gravity and dark energy models on galaxy clustering by computing the velocity-to-matter density ratio $\mathcal{F}$, the velocity contribution $\mathcal{R}$, and the potential contribution $\mathcal{P}$ and give an estimate of their detectability in future galaxy surveys.
Our results show that, in general, the relativistic correction contains additional information on gravity and dark energy, which needs to be taken into account in consistent horizon-scale tests of departures from $\Lambda$CDM using the galaxy-density field.

\end{abstract}

\maketitle


\section{Introduction}\label{sec:intro}

Modifications of gravity and dark energy models can serve as an alternative explanation to the cosmological constant causing the late-time accelerated expansion of our Universe.
There exists a large number of models that yield cosmic acceleration, and it is, therefore, of great interest to cultivate a formalism within which such models can systematically be explored while providing a consistent description of their small- and large-scale structure. 
Whereas the background evolution can be nearly identical for all models, their perturbations generally differ.
In `analogy' to the parametrized post-Newtonian formalism on local scales~\cite{will:05}, in the last few years, a parametrized post-Friedmann (PPF) framework~\footnote{The term `PPF' has been used by different authors. Notably~\cite{baker:11,baker:12} directly parametrize the contributions of spin-0 degrees of freedom from new scalar, vector, and tensor fields to the perturbed Einstein equation. This differs from the formalism of~\cite{PPF:07,PPF:08}, which introduces an effective fluid that reproduces the phenomenology of the corresponding modified gravity model at the perturbation level.} has been developed for the generalized computation of the linear perturbations around the spatially homogeneous and isotropic background of our Universe~\cite{uzan:06,caldwell:07,zhang:07,amendola:07,PPF:07,amin:07,bertschinger:08,PPF:08,pogosian:10,bean:10,hojjati:11,baker:11,baker:12,linder:05,koivisto:05,diporto:07,linder:09}.
The PPF formalism consistently unifies this computation for a large class of modified gravity and dark energy models on all linear scales, i.e., if correctly accounting for near-horizon modifications, and can be used to consistently test these models with structures from the largest observable scales.
Hence, it provides the proper generalized framework to study effects on the relativistic description of the galaxy-density field for modified gravity and dark energy cosmologies.

Among the most extensively studied models of modified gravity are scalar-tensor theories.
Such models appear naturally in attempts to unify general relativity (GR) with the standard model interactions.
In these models a scalar field appears in addition to the gravitational tensor field.
This fifth element or quintessence field~\cite{caldwell:97} may not couple directly to the matter fields but can source the cosmic acceleration observed today.
Thus, such \emph{minimally} coupled scalar-tensor models are natural candidates for dark energy.
The scalar field may, however, couple to the matter fields \emph{nonminimally}, yielding modifications of the gravitational interactions.
Such extended or generalized quintessence models have intriguing properties and may, e.g., act as the source of a phantom energy, where the equation of state associated with the scalar field becomes of the form $p < -\rho$.
Nonminimally coupled scalar-tensor theories are strongly limited by local measurements~\cite{will:05}.
However, nonlinear mechanisms that shield the extra force in such high-density environments and alleviate the constraints may appear~\cite{khoury:03}.
In the light of known (see also~\cite{vainshtein:72}) and possibly yet unknown suppression effects, it is important to independently test gravity on large scales in order to confirm the extrapolation of GR from the well-tested local region to the largest scales of our Universe.

With the advent of large-scale galaxy redshift surveys in the past few decades, galaxy clustering has become one of the most powerful measures of the large-scale inhomogeneity and probes GR on scales untested before.
In particular, in~\cite{samushia:12}, measurements of galaxy clustering in redshift space from the Baryonic Oscillation Spectroscopic Survey (BOSS) have been used to constrain deviations from GR parametrized by the logarithmic growth rate of structure $f=\Om^\gamma$.
Furthermore, constraints on the redshift evolution of $f$ are inferred in~\cite{blake:12} using data from the WiggleZ Dark Energy Survey.
All of their measurements are well described by a $\Lambda$CDM universe with GR.

It is important to keep in mind, however, that on larger scales, close to the horizon, where modified gravity or dark energy may drive the cosmic acceleration today, the standard Newtonian description of galaxy clustering becomes inaccurate as the relativistic effects become substantial.
Measurements on these large scales may be misinterpreted as the breakdown of GR on horizon scales.
Moreover, certain classes of models such as Dvali-Gabadadze-Porrati (DGP)~\cite{dvali:00} braneworld gravity possess a characteristic scale at which the graviton starts to propagate into the codimensions of the bulk embedding our 4D brane universe and features a transition from the effective gravity model in 4D to its full nature of modified gravity.
Therefore, testing GR against those models requires a proper covariant description of galaxy clustering and its measurements beyond the transition scale.

In the past few years, the relativistic description of galaxy clustering has been developed~\cite{yoo:09,yoo:10} to take into account the relativistic effects in galaxy clustering at large scales.
It is shown~\cite{yoo:09,yoo:10,bonvin:11,challinor:11,bruni:11,baldauf:11,jeong:11,bertacca:12} that on large scales, the signature of galaxy clustering in GR is significantly different from its Newtonian description and that the relativistic effects provide a new horizon-scale test of GR~\cite{yoo:12}.
The relativistic formula differs from the Newtonian description because our measurements of galaxy clustering are constructed by observing photons, and they are directly affected by the two metric potentials.
Compared to $\Lambda$CDM, in modified gravity and dark energy models, the metric potentials and the velocity respond differently to the same matter distribution.
This difference characterizing the modifications to $\Lambda$CDM may be prominent on very large scales, providing a unique opportunity to probe gravity and dark energy.

In this paper, we explore the observable signatures of modified gravity and dynamical dark energy in galaxy clustering.
We adopt the rather phenomenological but simple and efficient PPF formalism of~\cite{PPF:07} for a unified and consistent description of the associated linear perturbations.
For generalized scalar-tensor theories, horizon-scale PPF fits within the formalism of~\cite{PPF:07} have not been developed previously (see, however,~\cite{baker:12} for a description of scalar-tensor model perturbations in a different PPF approach).
Therefore, in order to consistently compute the near-horizon perturbations, we introduce a calibration method for these models.
We test our fits against the full model-specific linear perturbations and compare our results to existing PPF fits in the limit of $f(R)$ gravity, corresponding to a specific case of scalar-tensor gravity.
We then generalize the relativistic formula of galaxy clustering for the description of a PPF universe and explore the signature of scalar-tensor gravity and quintessence models in galaxy clustering, estimating their detection significance in future galaxy surveys.
In addition to the scalar-tensor models, we give predictions for the \emph{self-accelerating} and \emph{normal} branch DGP models as well as for phenomenological modifications of gravity, all described through the PPF formalism.

The organization of the paper is as follows.
In~\textsection\ref{sec:st}, we first present specific examples of minimally and nonminimally coupled scalar-tensor gravity models, referring to Appendix~\ref{app:sttheory} for the most important aspects of the underlying theory, including the background and perturbation evolutions.
We devote~\textsection\ref{sec:DGP} to the DGP braneworld model and~\textsection\ref{sec:pheno} to the phenomenological modifications of gravity.
In~\textsection\ref{sec:ppf}, we review the PPF formalism of~\cite{PPF:07} describing the linear perturbations on the cosmological background within these models.
In~\textsection\ref{sec:PPFST}, we describe our calibration method for computing the near-horizon perturbations in scalar-tensor models and discuss its performance in comparison to the full perturbation analysis.
The phenomenological modifications of gravity are defined through the PPF formalism in~\textsection\ref{sec:ppftestmod}.
We refer to Appendix~\ref{app:PPFfits} for the PPF fitting functions and parameters for $f(R)$ gravity~\cite{PPF:07} and DGP~\cite{PPF:07,lombriser:09}.
In~\textsection\ref{sec:observables}, we discuss the application of the PPF formalism to determine the signature of modified gravity and dark energy models in galaxy clustering.
We quantify their impact on galaxy clustering in~\textsection\ref{sec:relcorr} and present approximate estimates of their detectability in future galaxy surveys in~\textsection\ref{sec:significance}.
We conclude in~\textsection\ref{sec:conclusion} and refer to Appendix~\ref{app:sttheory} and~\ref{app:PPFfits} for the details on the codes employed for the numerical computations.

Throughout the paper, we assume a spatially homogeneous and isotropic universe.
Hence, the background metric is defined by the Friedmann-Lema\^itre-Robertson-Walker (FLRW) line element
\bq
 \rmd s^2 = - \rmd t^2 + a^2(t) \, \rmd \bf{x}^2, \label{eq:flrw}
\eq
where we further assume spatial flatness $K=0$.
We adopt $h=0.73$ and $\Om=0.24$ based on the $\Lambda$CDM cosmology inferred from the Wilkinson Microwave Anisotropy Probe~\cite{WMAP:10} in all modified gravity and dark energy models throughout the paper to highlight deviations from a $\Lambda$CDM universe.


\section{Dark energy and modified gravity}\label{sec:DEmodels}

In order to assess the effects of modified gravity and dark energy models on the relativistic description of galaxy clustering, we specialize to a number of well-studied quintessence and modified gravity models along with phenomenological modifications defined through the PPF formalism.
We summarize the models studied in this paper in Table~\ref{tab:models} (see~\textsection\ref{sec:significance}).
In this section, we shall give a brief model description, while the background and perturbation equations are described in~\textsection\ref{sec:ppf} as well as Appendix~\ref{app:sttheory} and~\ref{app:PPFfits}.
We devote~\textsection\ref{sec:st} to the minimally and nonminimally coupled scalar-tensor theory.
In~\textsection\ref{sec:DGP} and~\textsection\ref{sec:pheno}, we discuss the DGP braneworld scenario and phenomenological modifications of gravity, respectively.


\subsection{Scalar-tensor gravity} \label{sec:st}

We consider scalar-tensor theories in which the modified Einstein-Hilbert action in the Jordan frame is of the form
\bqa
 S & = & \frac{1}{2\kappa^2} \int d^4x \sqrt{-g} \left[ F(\scal) R - Z(\scal) \partial^{\mu} \scal \partial_{\mu} \scal - 2 U(\scal) \right] \nonumber \\
 & & + \Sm[\psi_{\rm m}; g_{\mu\nu}],
 \label{eq:action}
\eqa
where $\Sm$ is the matter action with matter fields $\psi_{\rm m}$, $\kappa^2 \equiv 8 \pi \, G$ with the bare gravitational coupling $G$, and we have set the speed of light $c\equiv1$.
Here, $F(\scal)$ describes the coupling of the scalar field $\scal$ to the metric, $Z(\scal)$ is the kinetic coupling, and $U(\scal)$ is the scalar-field potential.
In the Jordan frame action, Eq.~(\ref{eq:action}), matter fields are not explicitly coupled to the scalar field, hence, satisfying the weak equivalence principle.
Conformally transforming the metric $g_{\mu\nu}$ and redefining the scalar field $\scal$, the geometric part of the action and, subsequently, the Einstein field equations can be brought into their general relativistic form.
In this Einstein frame, matter fields are explicitly coupled to the new scalar field.
If $F={\rm const}$, the two metrics are equal up to the constant factor $F$, in which case matter fields are \emph{minimally} coupled.
Models with nonconstant $F$ are consequently referred to as \emph{nonminimally} coupled scenarios.
In this paper, we adopt the Jordan frame description.

With the freedom to redefine the scalar field $\scal$ in~Eq.~(\ref{eq:action}), reducing the number of free functions of $\scal$ to two instead of three,
we parametrize the nonminimally coupled models using the Brans-Dicke~\cite{brans:61} representation
\bq
 F\equiv\scal, \ \ \ \ \ Z\equiv\frac{\omega(\scal)}{\scal}. \label{eq:bdparam}
\eq
For the minimally coupled models, we set $F(\scal)=Z(\scal)\equiv1$.
We refer to~\cite{espositofarese:00} for the more general description of the background and perturbation equations using $F(\scal)$, $U(\scal)$, and $Z(\scal)$.
All of the modified gravity and dark energy models discussed in this paper, except for DGP and possibly the phenomenological modifications of gravity (see~\textsection\ref{sec:pheno}), can be formulated by an action in form of Eq.~(\ref{eq:action}).


\subsubsection{Minimally coupled examples} \label{sec:quintessence}

\begin{figure}
 \resizebox{\hsize}{!}{\includegraphics{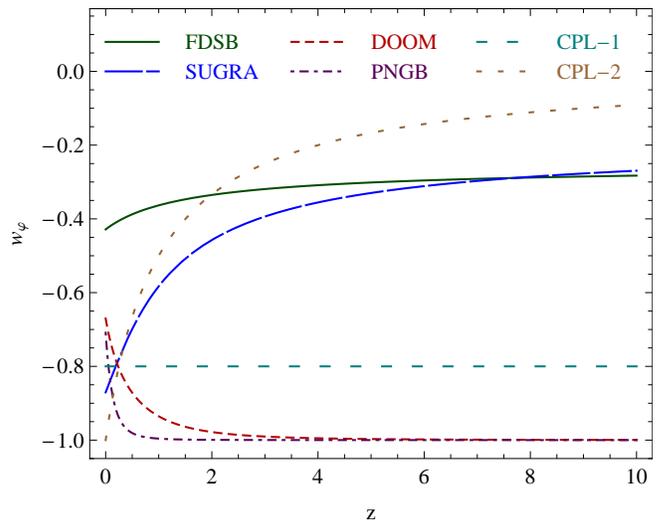}}
\caption{Dark energy equation of state associated with the quintessence models in~\textsection\ref{sec:quintessence}: the freezing models motivated by the fermion condensate model with dynamical supersymmetry breaking (FDSB) and by supergravity (SUGRA), where the scalar field rolls along the potential at early times but then slows down at late times; the thawing models with collapse at doomsday (DOOM) and with a scalar-field potential motivated by the pseudo-Nambu-Goldstone Boson (PNGB), where the scalar field is frozen at $w_{\scal}\simeq-1$ at early times and begins to evolve at later times; and the phenomenological quintessence models with constant and time-dependent dark energy equation of state.
}
\label{fig:wphi}
\end{figure}

We describe minimally coupled scalar-tensor models by the modified Einstein-Hilbert action Eq.~(\ref{eq:action}) with $F(\scal)=Z(\scal)\equiv1$.
We refer to Appenix~\ref{sec:QSpert} for the Einstein and conservation equations as well as the background equations inferred from them.
Note that minimally coupled scalar-tensor models do not modify gravity since the Einstein equations for the Jordan frame metric are not modified, i.e., the Einstein tensor relates to the energy-momentum tensor as in GR and the scalar field can be interpreted as an effective fluid contribution to the matter and radiation components with equation of state
\bq
 w_{\scal} \equiv \frac{p_{\scal}}{\rho_{\scal}} = \frac{(H\,\scal')^2 - 2 U}{(H\,\scal')^2 + 2 U}, \label{eq:quintwscal}
\eq
where here and throughout the paper, primes denote derivatives with respect to $\ln a$.

In this paper, we study four different analytic potentials of $\scal$~\cite{tsujikawa:10} and, in addition, two potentials phenomenologically defined by an effective Hubble expansion.
We show the scalar-field equation of state for the different models in Fig.~\ref{fig:wphi}.

\paragraph{Freezing models:}

As our first examples of quintessence potentials, we consider two \emph{freezing} models, in which the field rolls along the potential at early times but then slows down at late times,
\bqa
 {\rm FDSB:} & & U(\scal) = \frac{M_0^{4+n}}{\scal^n}, \ \ \ \ n>0, \label{eq:fdsb} \\
 {\rm SUGRA:} & & U(\scal) = \frac{M_0^{4+n}}{\scal^n} e^{\alpha \, \scal^2}, \label{eq:sugra}
\eqa
where $M_0$, $\alpha$, and $n$ are the model-specific parameters.
The potential in Eq.~(\ref{eq:fdsb}) is motivated by the fermion condensate model with a dynamical supersymmetry breaking (FDSB)~\cite{ratra:87, binetruy:98, steinhardt:99} and
the potential in Eq.~(\ref{eq:sugra}) may arise from supergravity (SUGRA) models~\cite{brax:99}.
We set the FDSB potential with $n=6$ in Eq.~(\ref{eq:fdsb}) and the SUGRA potential with $n=11$ and $\alpha=1/2$ in Eq.~(\ref{eq:sugra}).
The scalar field $\scal$ at early epoch and the characteristic mass $M_0$ are adjusted to match the present-day Hubble parameter $H^2(a=1)=H_0^2$.

\paragraph{Thawing models:}

The next two potentials of interest here are classified as \emph{thawing} models. In these scenarios, the scalar field is frozen at early times with $w_{\scal}\simeq-1$ and begins to evolve at late times,
\bqa
 {\rm DOOM:} & & U(\scal) = V_0 + M_0^{4-n} \scal^n, \ \ \ \ n>0 \label{eq:doom} \\
 {\rm PNGB:} & & U(\scal) = M_0^4 \cos^2\left(\alpha \, \scal \right). \label{eq:pngb}
\eqa
The potential in Eq.~(\ref{eq:doom}) was studied in~\cite{kallosh:03} as a model in which $\scal$ rolls down the potential and collapses at \emph{doomsday} (DOOM).
The potential in Eq.~(\ref{eq:pngb}) is motivated assuming the presence of an ultralight pseudo-Nambu-Goldstone Boson (PNGB)~\cite{frieman:95}.
We set the DOOM potential with $n=1$ and $V_0=3H_0^2$ in Eq.~(\ref{eq:doom}) and the PNGB potential with $\alpha=0.445^{-1}$ in Eq.~(\ref{eq:pngb}).
The characteristic mass $M_0$ and the initial conditions for $\scal$ are again adjusted to match $H^2(a=1)=H_0^2$.

\paragraph{Phenomenological quintessence models:} \label{sec:phenoquintessence}

The last two quintessence models we analyze here are purely phenomenological and are based on the Chevallier-Polarski-Linder (CPL)~\cite{chevallier:00,linder:02} parametrization of the dark energy equation of state $w_{\scal}=w_0+(1-a)w_a$.
In particular, we consider the two cases
\bqa
 {\rm CPL-1:} & & w_{\scal} = w_0, \label{eq:phenquint1} \\
 {\rm CPL-2:} & & w_{\scal} = -a. \label{eq:phenquint2}
\eqa
Note that an equation of state of the form $w_{\scal}\approx-a$ can be made compatible with current cosmological observations by introducing anisotropic stress~\cite{lombriser:11a}.
Here, we shall, however, restrict to cases with no anisotropic stress (see, however,~\textsection\ref{sec:ppftestmod}), which is a natural consequence of the minimal coupling with a constant $F$ (see~\textsection\ref{sec:STPPF}).
For the reconstruction of the scalar field, we treat its contribution to the expansion history as an effective fluid with equation of state $w_{\scal}$ and use Eqs.~(\ref{eq:quintwscal}) and (\ref{eq:quintFriedmann1}) to determine $U$ and $\scal'$.
We illustrate the dark energy equations of state of Eqs.~(\ref{eq:phenquint1}) and (\ref{eq:phenquint2}) in Fig.~\ref{fig:wphi}, where $w_0=-0.8$ in Eq.~(\ref{eq:phenquint1}).


\subsubsection{Nonminimally coupled examples} \label{sec:nonmincoupexmp}

We describe nonminimally coupled scalar-tensor models in the modified Einstein-Hilbert action, Eq.~(\ref{eq:action}), in their Brans-Dicke representation, Eq.~(\ref{eq:bdparam}).
We refer to Appenix~\ref{sec:BDpert} for the modified Einstein and scalar-field equations as well as the background equations inferred from them.
As representative examples, we study two models of nonminimally coupled scalar-tensor theories: $f(R)$ gravity~\cite{buchdahl:70,starobinsky:80,carroll:03,nojiri:03,capozziello:03} or $\omega=0$ Brans-Dicke models, i.e., with vanishing kinetic term $Z=0$ in the Jordan frame action, Eq.~(\ref{eq:action}), supplied with a scalar-field potential, and a scalar-tensor model with nonconstant Brans-Dicke parameter~\cite{sanchez:10}.

\begin{figure}
 \resizebox{\hsize}{!}{\includegraphics{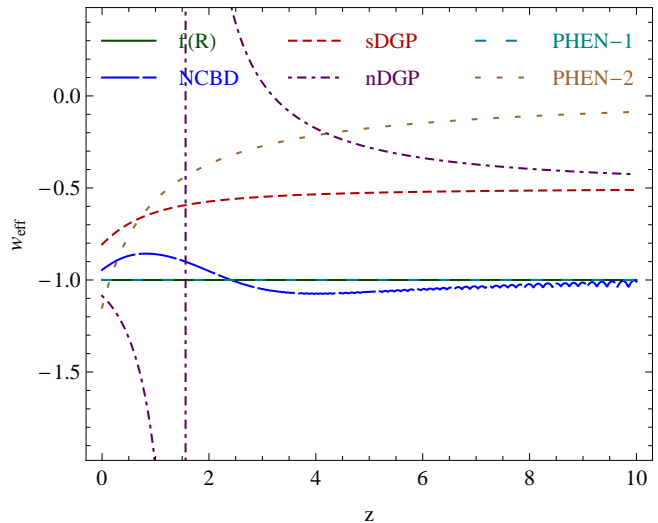}}
\caption{Effective dark energy equation of state associated with different modified gravity models.
As in Fig.~\ref{fig:wphi}, we consider six example models: two nonminimally coupled scalar-tensor theories, $f(R)$ gravity and the nonconstant Brans-Dicke (NCBD) parameter model; the two branches of the DGP model, the self-accelerating (sDGP) and normal (nDGP) branch; and two phenomenological modifications (PHEN-1 and PHEN-2). By design, PHEN-1 and the $f(R)$ models considered here recover the background expansion history of a $\Lambda$CDM universe, where $w=-1$. Hence, their equations of state overlap.
}
\label{fig:weff}
\end{figure}

\paragraph{$f(R)$ gravity:}\label{sec:fRgravity}

The action of metric $f(R)$ gravity models is obtained from Eqs.~(\ref{eq:action}) and~(\ref{eq:bdparam}) by defining
\bqa
 \scal & \equiv & 1 + f_R, \label{eq:fRreplace1} \\
 U & \equiv & \frac{R \, f_R - f}{2}, \label{eq:fRreplace2} \\
  \omega & \equiv & 0, \label{eq:fRreplace3}
\eqa
where $f(R)$ is a nonlinear function of the Ricci scalar and $f_R\equiv \rmd f(R)/\rmd R$.
We focus here on designer $f(R)$ models~\cite{song:06}, where $f(R)$ is reconstructed from a given Hubble expansion.
We shall require a strict matter-dominated $\Lambda$CDM expansion history, hence, we set $H^2=\kappa^2(\rhm+\rho_{\Lambda})/3$ in the $f(R)$ Friedmann equation,
which is obtained by applying the replacements of Eqs.~(\ref{eq:fRreplace1}), (\ref{eq:fRreplace2}), and (\ref{eq:fRreplace3}) to Eq.~(\ref{eq:BDfriedmann1}).
The Friedmann equation then becomes an inhomogeneous second-order differential equation for $f(R)$ as a function of $\ln a$,
\bq
 f'' - \left[ 1 + \frac{H''}{H} + \frac{R''}{R'} \right] f' + \frac{R'}{6 H^2} f = -H_0^2 (1-\Omega_{\rm m}) \frac{R'}{H^2}.
\eq
This relation can be solved numerically with the initial conditions
\bqa
 f(\ln a_{\rm i}) = A \, H_0^2 a_{\rm i}^p - 6 H_0^2 (1-\Om), \\
 f'(\ln a_{\rm i}) = p \, A \, H_0^2 a_{\rm i}^p,
\eqa
where $p = (-7 + \sqrt{73})/4$.
$A$ is an initial growing mode amplitude and characterizes a specific solution in the set of functions $f(R)$ that recover the $\Lambda$CDM background.
The amplitude of the decaying mode is set to zero in order to not violate high-curvature constraints.
Instead of defining the solutions by setting $A$ at an arbitrary redshift, we define them by a condition set today.
This can be done by, e.g., either characterizing them at the background via $f_{R0} \equiv f_R(a=1)$ or via the Compton wavelength parameter
\bq
 B = \frac{f_{RR}}{1+f_R} R' \frac{H}{H'} \label{eq:compton}
\eq
evaluated today, $B_0 \equiv B(a = 1)$.
GR is recovered in the case where $B_0=0$ or, equivalently, $f_{R0}=0$.

Currently, constraints on $f(R)$ gravity are of the order $B_0\lesssim10^{-3}$ or $|f_{R0}|\lesssim10^{-4}$ (95\% confidence level) from cosmological probes~\cite{schmidt:09c,lombriser:10,lombriser:11b}, $|f_{R0}|\lesssim10^{-6}$ from Solar System constraints~\cite{hu:07}, corresponding to the requirement of a chameleon-screened Milky Way halo~\cite{hu:07,brax:08,lombriser:12}, and $|f_{R0}|\lesssim10^{-7}$ from distance measurements and stellar physics~\cite{jain:12}.
An upper limit of $|f_{R0}|\lesssim10^{-5}$ is expected from future 21~cm survey data when combined with cosmic microwave background (CMB) data from {\it Planck}~\cite{hall:12}.
Also see~\cite{hu:12} for expected constraints on $f(R)$ gravity from a joint analysis of the bispectrum of the correlation of the integrated Sachs-Wolfe (ISW) effect with weak gravitational lensing and the CMB power spectrum from {\it Planck} data.
Despite the strong constraints on $f(R)$ gravity, the model serves as a useful toy model for the study of the large-scale structure in modified gravity.
We show the effective dark energy equations of state for the $f(R)$ gravity models studied here in Fig.~\ref{fig:weff}.
The models are designed to exactly match the $\Lambda$CDM expansion history, hence, $w_{\rm eff}=-1$.

\paragraph{Nonconstant Brans-Dicke parameter model:}\label{sec:NCBD}

Our second example with a nonminimal coupling is equivalent to a scalar-tensor model studied in~\cite{sanchez:10}.
It has a nonconstant Brans-Dicke (NCBD) parameter $\omega(\scal)$
and is defined by
\bqa
 U & \equiv & H_0^2 \left( 1 + e^{u_0\sqrt{1-\scal}} \right), \label{eq:NCBD1} \\
 \omega & \equiv & \omega_0 \frac{\scal}{1-\scal}, \label{eq:NCBD2} 
\eqa
where we set the two model parameters $\omega_0=1/40$ and $u_0=-3$ to emphasize near-horizon modified gravity effects on the effective anisotropic stress in~\textsection\ref{sec:PPFperformance}.
For more details on the numerical integration, we refer to~\cite{sanchez:10} or Appendix~\ref{sec:NCBDbackground}.
We illustrate the effective dark energy equation of state of the model in Fig.~\ref{fig:weff}, which shows a phantom crossing with $w_{\rm eff}\leq-1$ and $w_{\rm eff}>-1$ at early and late times, respectively.


\subsection{DGP braneworld gravity}\label{sec:DGP}

We extend our analysis of modified gravity to the well-studied DGP braneworld model~\cite{dvali:00}.
Note that since the PPF fits for the two DGP branches have been developed in~\cite{PPF:07,lombriser:09,seahra:10,koyama:05}, we only provide a short schematic review and refrain from describing the modified Einstein equations and their perturbations within DGP in the their full 5D form (see, however,~\cite{deffayet:00,deffayet:02,koyama:05,sawicki:06,song:07b,cardoso:07,seahra:10}).

In the DGP model, our Universe is a $(3+1)$-brane embedded in a 5D Minkowski space described by the action
\bqa
 S & = & -\frac{1}{2\kappa^2} \int d^5 x \sqrt{-\hat{g}}{\hat{R}} - \frac{1}{2\mu^2} \int d^4x \sqrt{-\tilde{g}}{\tilde{R}} \nonumber \\
 & & + \int d^4x \sqrt{-\tilde{g}} L_{\rm T}, \label{eq:DGPaction}
\eqa
where 5D quantities are denoted by hats and 4D quantities are denoted by tildes. Matter fields, including a cosmological constant or brane tension, are represented by $L_{\rm T}$ and confined to the brane, while only gravity can propagate through the full 5D bulk. We assume that there is no bulk tension.
The constants $\kappa^2$ and $\mu^2$ are proportional to the inverse Planck masses in the bulk and brane, respectively.
Gravity on the brane is consequently modified at large scales.
In particular, the crossover distance $r_{\rm c} = \kappa^2/2\mu^2$ governs the transition from 5D to 4D scalar-tensor gravity and determines the nonlinear screening, e.g., on scales smaller than the Vainshtein~\cite{vainshtein:72} radius $r_* = (r_{\rm c}^2 r_g)^{1/3}$, nonlinear interactions return gravity to GR around a point mass with Schwarzschild radius $r_g$. 

Variation of the action Eq.~(\ref{eq:DGPaction}) yields the modified Einstein equations on the brane, which for a homogeneous and isotropic metric reduce to the modified Friedmann equation~\cite{deffayet:00}
\bq
 H^2 - \sigma \frac{H}{r_{\rm c}} = \frac{\mu^2}{3} \sum_{\rm i} \rho_{\rm i},
\eq
where $\rho_{\rm i}$ are the energy densities of various components on the brane, and we assume vanishing spatial curvature on the brane.
The sign $\sigma=\pm1$ defines the branch of the cosmological solutions. For $\sigma=+1$, late-time acceleration occurs even without a cosmological constant $\Lambda$~\cite{deffayet:00}, and so this branch is referred to as self-accelerating DGP (sDGP). For $\sigma=-1$, DGP modifications slow the expansion rate, and the branch is referred to as normal branch (nDGP). Here, a cosmological constant or additional dark energy is required to achieve late-time acceleration.
For a spatially flat matter-only universe with possible contribution of a brane tension or cosmological constant $\Lambda$, the Friedmann equation becomes
\bq
 H = H_0 \left(\sqrt{ \Om a^{-3} + \Omega_{\Lambda} + \Omega_{r_{\rm c}}} + \sigma \sqrt{\Omega_{r_{\rm c}}} \right), \label{eq:dgpfriedmann}
\eq
with the density parameter associated with the crossover distance
\bq
 \sqrt{\Omega_{r_{\rm c}}} \equiv \frac{1}{2H_0 r_{\rm c}} = \sigma \frac{1-\Om-\Omega_{\Lambda}}{2}, \label{eq:Omegarc}
\eq
where the second equality follows from the condition $H(a=1)=H_0$.

The DGP equation of state is given by
\bq
 1 + w_{\rm DGP} = \frac{\mu^2\sum_{\rm i} (1+w_{\rm i}) \rho_{\rm i}}{3H^2 + \mu^2\sum_{\rm i} \rho_{\rm i}},
\eq
where for cases with a cosmological constant, it is also useful to define the total effective dark energy $\rho_{\rm eff} = \rho_{\rm DGP} + \rho_{\rm \Lambda}$ and its equation of state
\bq
 1 + w_{\rm eff} = (1+w_{\rm DGP}) \frac{\rho_{\rm DGP}}{\rho_{\rm DGP} + \rho_{\Lambda}}
\eq
with $\rho_{\rm DGP} \equiv 3H^2/\mu^2 - \sum_{\rm i} \rho_{\rm i}$.
We show the effective dark energy equation of state associated with the DGP modifications in Fig.~\ref{fig:weff}, where we use the fiducial cosmological parameters and $\Omega_{\Lambda}=0$ or $\Omega_{\Lambda}=1.51$ to determine the crossover distance in Eq.~(\ref{eq:Omegarc}) for the sDGP or nDGP models, respectively.
We use these parameter values throughout the paper.
The effective dark energy equation of state in the nDGP scenario has a singularity at $z\sim1.5$ where it jumps across the phantom divide with $w_{\rm eff}>-1$ and $w_{\rm eff}<-1$ at early and late times, respectively.
Note that the DGP background is completely defined by $w_{\rm eff}$ and energy-momentum conservation of the brane components and the energy-density of the effective fluid $\rho_{\rm eff}$.

The sDGP model is in conflict with cosmological data and can be ruled out at more than $5\sigma$ for both flat and nonflat universes~\cite{fang:08a,lombriser:09}.
If allowing for a brane tension in this branch (sDGP+$\Lambda$), a crossover scale of $H_0 r_{\rm c}\gtrsim3$ (95\% confidence level) is required for compatibility with cosmological observations, which essentially reduces it to a $\Lambda$CDM universe~\cite{lombriser:09}.
A bound of the same order can be found for flat and nonflat nDGP universes, where in the nonflat case, the cross-correlation of the CMB temperature anisotropy with foreground galaxies through the ISW effect are used to break a degeneracy between the spatial curvature and the nDGP modification in the CMB temperature power spectrum and expansion history~\cite{lombriser:09,giannantonio:08}.
Recently, constraints on a nDGP model with $\Lambda$CDM expansion history, i.e., with an appropriate additional dark energy contribution, have been obtained from redshift-space distortions~\cite{raccanelli:12}.
Despite the strong constraints and theoretical difficulties of the self-accelerating solution~\cite{koyama:07}, the DGP braneworld model still serves as a useful toy model for the study of the large-scale structure in modified gravity.


\subsection{Phenomenological modifications of gravity}\label{sec:pheno}

We can study deviations of GR in a more general context and allow for phenomenological modifications of gravity, where effective modifications in the Hubble expansion as well as in the linear perturbations are introduced without a direct representation through a geometric modification of the Einstein-Hilbert action or Einstein equations but are formulated by an effective fluid contribution.
Due to the energy-momentum conservation of the matter component and the Bianchi identities, the energy-momentum tensor associated with this fluid is also covariantly conserved.
While the linear perturbations of the fluid are characterized via the PPF formalism in~\textsection\ref{sec:ppftestmod}, at the background level, we describe the modification directly through its effective dark energy equation of state $w_{\rm eff}(a)$ [cf.~Eq.~(\ref{eq:enmomeff})].
The expansion history is then given by
\bqa
 H^2 & = & \frac{\kappa^2}{3}(\rhm + \rho_{\rm eff}), \label{eq:phenofriedmann} \\
 \rho_{\rm eff}' & = & -3(1+w_{\rm eff})\rho_{\rm eff}, \label{eq:phenoconservation}
\eqa
where we can write the dark energy density associated with the modification as
\bqa
 \rho_{\rm eff} & = & \rho_{{\rm eff}0} a^{-3(1+\tilde{w}_{\rm eff})} \nonumber\\
 & = & \rho_{{\rm eff}0} \exp \left[ 3 \int_a^1 \frac{1+w_{\rm eff}(a')}{a'} \rmd a' \right]. \label{eq:weff}
\eqa
Here, we implement the CPL parametrization, which we have also used in~\textsection\ref{sec:phenoquintessence},
\bq
 w_{\rm eff}(a) = w_0 + (1-a) w_a. \label{eq:phenoback}
\eq
In this case, Eq.~(\ref{eq:weff}) simplifies to
\bq
 \rho_{\rm eff} = \rho_{{\rm eff}0} a^{-3(1+w_0+w_a)} \exp[3w_a(a-1)]. \label{eq:rhomod}
\eq

We shall consider two specific cases of phenomenological (PHEN) modifications,
\bqa
 {\rm PHEN-1:} & & \ \ w_0=-1, \ \ \ \ \ \ \ w_a=0, \\
 {\rm PHEN-2:} & & \ \ w_0=-1.15, \ \ \ w_a=1.17.
\eqa
The former follows a $\Lambda$CDM expansion history, and the latter is motivated by the best-fit parameters to current cosmological observations found in~\cite{lombriser:11a} (see~\textsection\ref{sec:ppftestmod}).
At the background level, the phenomenological modification of gravity is equivalent to a corresponding phenomenological quintessence model and completely defined by the effective dark energy equation of state, Eq.~(\ref{eq:phenoback}), along with the Friedmann equation, Eq.~(\ref{eq:phenofriedmann}), and energy-momentum conservation, Eq.~(\ref{eq:phenoconservation}).
In particular, PHEN-1 is equivalent to a $\Lambda$CDM model.
They will only be completely characterized and distinguishable at the perturbation level, which shall be elucidated in~\textsection\ref{sec:ppftestmod}.
Note that without a proper consideration of perturbations, the time-evolving dark energy models are gauge-dependent and theoretically inconsistent.

We show the effective dark energy equations of state $w_{\rm eff}$ for our phenomenological modifications in Fig.~\ref{fig:weff}.
The first phenomenological modified gravity model expands equivalently to $\Lambda$CDM, overlapping with the $w$ of the designer $f(R)$ model.
The second phenomenological modified gravity model has a phantom crossing with $w_{\rm eff}<-1$ at late times and approaches a matterlike equation of state $w_{\rm eff}\simeq0$ at early times.


\section{Parametrized Post-Friedmann framework} \label{sec:ppf}

In `analogy' to the parametrized post-Newtonian formalism in the Solar System~\cite{will:05}, the PPF framework (see, e.g.,~\cite{PPF:07,hojjati:11,baker:11,baker:12}) provides a unified description of the linear perturbation theory around the FLRW background for generalized modified gravity and dark energy models.
In this framework, the extra terms appearing in the Einstein field equations due to scalar-field contributions or modifications of gravity, e.g., Eqs.~(\ref{eq:QSEinstein}) and~(\ref{eq:BDEinstein}), are viewed as an effective dark energy fluid component, defined as
\bq
 T_{\rm eff}^{\mu\nu} \equiv \kappa^{-2} G^{\mu\nu} - T_{\rm m}^{\mu\nu}, \label{eq:enmomeff}
\eq
with energy-momentum conservation, $\nabla_{\mu} T_{\rm eff}^{\mu\nu} = 0$, due to the Bianchi identities and energy-momentum conservation of the matter components.
Thus, the usual cosmological perturbation theory may be applied with each four degrees of freedom in both the perturbation of the metric and of the energy-momentum tensor.
The Einstein and conservation equations fix four degrees of freedom, and the gauge choice fixes another two.
The remaining two degrees of freedom are then specified by two closure relations, which are defined by the particular modified gravity or dark energy model.
These two closure relations can be designed such that the effective fluid mimics the relations between the metric and matter perturbations given by the full perturbations of a particular modified gravity or dark energy cosmology.

This simple and generalized treatment allows for an efficient and consistent computation of the evolution of perturbations on large scales in modified gravity and dark energy theories.
Parametrizations of this kind can be applied to efficiently explore the parameter space of modified gravity and dark energy models employing computationally intensive generalized Boltzmann linear theory solvers and Markov chain Monte-Carlo techniques~\cite{daniel:10,pogosian:10,bean:10,lombriser:11a,hojjati:11,dossett:11b}.
A great advantage of the PPF formalism is that the corresponding numerical codes need only to be adapted once to employ the PPF modifications rather than for each nonstandard model separately~\cite{fang:08b,daniel:10,hojjati:11,dossett:11b}.
Such a generalized approach motivates the application of the PPF formalism to our study, providing the base upon which future horizon-scale tests of gravity and dark energy models employing, e.g., a multitracer analysis of galaxy-redshift survey data can be performed.

Developing a PPF formalism for general theories of modified gravity has been a subject of intensive study (see, e.g.,~\cite{uzan:06,caldwell:07,zhang:07,amendola:07,PPF:07,amin:07,bertschinger:08,PPF:08,pogosian:10,bean:10,hojjati:11,baker:11,baker:12,linder:05,koivisto:05,diporto:07,linder:09}).
In this paper, we follow~\cite{PPF:07}, adopting a phenomenological approach to the PPF formalism.
It provides an efficient general framework to account for linear perturbations of modified gravity theories, consistently describing the sub-, near-, and superhorizon scales.
Within this framework, parametrizations of $f(R)$ and DGP gravity have been developed in~\cite{PPF:07} and~\cite{PPF:07,lombriser:09,seahra:10}, respectively.
In the case of DGP and $f(R)$ models (with large values of $|f_{R0}|$), the importance of the consistent inclusion of near-horizon modifications for the accurate determination of the low multipoles of the CMB temperature fluctuations have been pointed out in~\cite{song:06b,song:07b,fang:08a,lombriser:09,lombriser:10} (cf.~\cite{hojjati:12}).
Hence, in order to get consistent CMB constraints on the DGP and $f(R)$ models, the DGP and $f(R)$ PPF functions, respectively, have been employed in parameter analyses of these models~\cite{fang:08a,lombriser:09,lombriser:10}.
The PPF formalism has also been used to consistently allow for phenomenological horizon-scale modifications of gravity and study constraints on such deviations from $\Lambda$CDM~\cite{lombriser:11a}.
Thereby, the simplicity of the PPF formalism described in~\cite{PPF:07} proved to be of great advantage in the efficient exploration of the parameter space through Boltzmann linear theory solvers~\cite{fang:08a,fang:08b} that have been employed to obtain accurate model constraints.

We briefly review the PPF framework of~\cite{PPF:07} in~\textsection\ref{sec:formal}.
In addition to the known PPF model descriptions in this formalism, in~\textsection\ref{sec:STPPF}, we describe a simple procedure to calibrate near-horizon corrections to the subhorizon PPF modifications for generalized minimally and nonminimally coupled scalar-tensor gravity models of the kind described by the modified Einstein-Hilbert action in Eq.~(\ref{eq:action}), disclosing consistent near-horizon tests of these models through the CMB or multitracer analyses of galaxy clustering.
We define the phenomenological modifications of gravity in~\textsection\ref{sec:ppftestmod} and discuss the PPF fitting functions for $f(R)$ gravity and the DGP model in Appendices~\ref{sec:fRPPF} and \ref{sec:ppf_dgp}, respectively.


\subsection{PPF formalism} \label{sec:formal}

We introduce the linear PPF perturbation formalism of~\cite{PPF:07} to compute linear perturbations of modified gravity and dark energy models.
Thereby, we follow the lead of~\cite{bardeen:80,kodama:85}.
For simplicity, we further restrict the linear perturbations to a spatially flat universe.
The formalism can, however, easily be extended to include radiation components and spatial curvature \cite{PPF:08}.
In the longitudinal gauge for the metric perturbations ($B=0 = H_T = 0$, $\Phi \equiv H_L$, $\Psi \equiv A$ in Bardeen's notation \cite{bardeen:80}), assuming that there is no anisotropic stress $\Pi_{\rm m} = 0$, the Einstein equations may be combined to yield
\bqa
 \Phi_+ & = & - \frac{\kappa^2}{2 H^2 k_H^2} p_{\rm eff} \Pi_{\rm eff}, \label{eq:PPFEinstein1} \\
 \Phi_- & = & \frac{\kappa^2}{2 H^2 k_H^2} \left[ \rho_{\rm m} \Delta_{\rm m} + \rho_{\rm eff} \Delta_{\rm eff} \right. \nonumber \\
  & & \left. + 3 (\rho_{\rm eff} + p_{\rm eff} ) \frac{V_{\rm eff} - V_{\rm m}}{k_H} + p_{\rm eff} \Pi_{\rm eff} \right], \label{eq:PPFEinstein2}
\eqa
where $\Phi_+ \equiv (\Phi + \Psi)/2$, $\Phi_- \equiv (\Phi - \Psi)/2$, $k_H \equiv k/(a \, H)$, and $w_{\rm eff} \equiv p_{\rm eff}/\rho_{\rm eff}$ is determined by the background relations.
Using the energy-momentum conservation of matter,
\bqa
\Delta'_{\rm m} & = & -k_H V_{\rm m} - 3~\zeta', \label{eq:emcons}\\
\vm' & = & k_H \Psi - \vm,
\eqa
and the above Einstein equations, we obtain the velocity of the effective fluid, 
\bq
 V_{\rm eff} = V_{\rm m} - k_H \frac{2H^2}{\kappa^2 a^2 (\rho_{\rm eff} + p_{\rm eff})} ~\zeta', \label{eq:veff}
\eq
and its pressure fluctuation, 
\bq
 \Delta p_{\rm eff} = p_{\rm eff} \Delta_{\rm eff} - \frac{1}{3} \rho_{\rm eff} \Delta'_{\rm eff} - (\rho_{\rm eff} + p_{\rm eff}) \left( \frac{k_H V_{\rm eff}}{3} + \zeta' \right),
\eq
where both quantities are in the longitudinal gauge and the comoving-gauge curvature is $\zeta=\Phi - {V_{\rm m}/ k_H}$.
These equations, in fact, define the effective fluid at the perturbation level.

However, modified gravity theories are not characterized by these properties of individual fluid components but are characterized by modifications of the Einstein equation.
In the PPF formalism of~\cite{PPF:07}, the Poisson equation is modified as
\bq
 k^2( \Phi_- + \Gamma) = \frac{\kappa^2}{2} a^2 \rho_{\rm m} \Delta_{\rm m}.
 \label{eq:poissoneq}
\eq
Together with the Einstein Eqs.~(\ref{eq:PPFEinstein1}) and~(\ref{eq:PPFEinstein2}), this yields the two PPF closure relations
\bqa
 -\frac{2k^2}{\kappa^2 a^2} \Gamma(a,k) & = & \rho_{\rm eff} \Delta_{\rm eff} + p_{\rm eff} \Pi_{\rm eff} \nonumber \\
 & & + 3(\rho_{\rm eff} + p_{\rm eff}) \frac{V_{\rm eff}-V_{\rm m}}{k_H}, \label{eq:closure_gamma} \\
  - \frac{2H^2 k_H^2}{\kappa^2} g(a,k) \Phi_- & = & p_{\rm eff} \Pi_{\rm eff}, \label{eq:closure_g}
\eqa
where the metric ratio is defined as
\bq
 g(a,k)\equiv\frac{\Phi_+}{\Phi_-} = \frac{\Phi+\Psi}{\Phi-\Psi}.
\label{eq:gPPFdef} 
\eq
Under the PPF framework, the linear perturbations of modified gravity theories and dark energy models are completely described by the two additional functions $\Gamma(a,k)$ and $g(a,k)$, each of which defines the modification of the Poisson equation in Eq.~(\ref{eq:closure_gamma}) and the anisotropic pressure of the effective fluid in Eq.~(\ref{eq:closure_g}), respectively (cf.~\cite{caldwell:07,zhang:07,amendola:07,bertschinger:08,hojjati:11,baker:11,baker:12}).
In terms of these two PPF functions, the conservation equations become
\bqa
 \dm' & = & -k_H \vm - 3(g+1) \phim' - 3 (1-g+g') \phim \nonumber \\
 & & + 3 \frac{H'}{H} \frac{\vm}{k_H}, \label{eq:PPFenergyconservation} \\
 \vm' & = & -\vm + (g-1) k_H \phim, \label{eq:PPFmomentumconservation}
\eqa
where we rewrite the metric perturbations in terms of $\phim$.

Given the metric ratio~$g(a,k)$ of each modified gravity model with its background evolution, the other PPF function~$\Gamma(a,k)$ is constrained in two limiting regimes, the super- and quasistatic subhorizon scales.
Since the energy-momentum is locally conserved even in modified gravity theories, the comoving gauge curvature is conserved on superhorizon scales~\cite{bertschinger:06,bardeen:80,PPF:07,hu:98,sawicki:06,song:06}
\bq
\zeta'=\mathcal{O}(k_H^2\zeta),
\label{eq:superhorizonconstraint}
\eq
and this relation constrains the PPF function~$\Gamma$ on horizon scales.
With Eqs.~(\ref{eq:poissoneq}), (\ref{eq:PPFenergyconservation}), and~(\ref{eq:PPFmomentumconservation}), the relation on superhorizon scales in Eq.~(\ref{eq:superhorizonconstraint}) can be used to yield the constraining equation on $\Gamma$ in the limit $k_H\rightarrow0$ as
\bq
\Gamma'+\Gamma=S,
\eq
where the source function is
\bqa
 S & = & - \left[ \frac{1}{g+1} \frac{H'}{H} + \frac{3}{2} \frac{H_0^2 \Omega_{\rm m}}{H^2 a^3} (1+f_{\zeta}) \right] \frac{V_{\rm m}}{k_H} \nonumber \\
 & & + \left[ \frac{g' - 2g}{g+1} \right] \Phi_-. \label{eq:Gammasource}
\eqa
In the source function~$S$, we parametrize the leading-order correction to $\zeta'$ as
\bq
 \lim_{k_H \rightarrow 0} {\zeta' \over k_H^2}
\equiv \frac{1}{3}~ f_{\zeta}~ {V_{\rm m}\over k_H}\sim\mathcal{O}(\zeta).
\eq
For an effective scalar-tensor theory where photon geodesics are not affected by the additional scalar degree of freedom, the other constraint on the PPF function~$\Gamma$ arises from the modified Poisson equation in the quasistatic regime on subhorizon scales
\bq
 k^2 \Phi_- = \frac{\kappa^2}{2(1+f_G)} ~a^2 \rho_{\rm m} \Delta_{\rm m}, \label{eq:qspoissoneq}
\eq
which implies $\Gamma\rightarrow f_G\phim$ in the limit $k_H\rightarrow\infty$, where the modification $f_G(a)$ is defined at the linear perturbation level, ignoring nonlinear suppression mechanisms.
Given the constraints in the two limiting regimes, a smooth interpolation of the PPF function $\Gamma(a,k)$ can be achieved by solving the evolution equation~\cite{PPF:07}
\bq
 (1 + c_{\Gamma}^2 k_H^2) [ \Gamma' + \Gamma + c_{\Gamma}^2 k_H^2 (\Gamma - f_G \Phi_-) ] = S, \label{eq:Gamma}
\eq
where the additional parameter $c_{\Gamma}$ relates the transition scale between the two limiting regimes to the Hubble scale.

Therefore, the PPF function~$\Gamma(a,k)$ is further parametrized by the transition scale~$c_\Gamma$ and two time-dependent functions $f_{\zeta}(a)$ and $f_G(a)$, each of which relates the matter to the metric on superhorizon scales and defines the modified Poisson equation in the Newtonian regime, respectively.
With the interpolation of Eq.~(\ref{eq:Gamma}), the PPF formalism completely defines the linear perturbations given the expansion history $H$ or $w_{\rm eff}$, along with the PPF functions and parameter, $g$, $f_\zeta$, $f_G$, and $c_{\Gamma}$.
These functions can be described for modified gravity theories, e.g., for DGP~\cite{PPF:07,lombriser:09,seahra:10} (see Appendix~\ref{sec:ppf_dgp}), $f(R)$ gravity~\cite{PPF:07} (see Appendix~\ref{sec:fRPPF}), and scalar-tensor models of the kind described by Eq.~(\ref{eq:action}) (see~\textsection\ref{sec:STPPF}), to obtain the linear fluctuations for the specific model.
In order to determine the PPF linear perturbations, we need to solve the coupled system of differential and constraint equations given by the conservation equations, Eqs.~(\ref{eq:PPFenergyconservation}) and (\ref{eq:PPFmomentumconservation}); the evolution equation for $\Gamma$, Eq.~(\ref{eq:Gamma}), with source Eq.~(\ref{eq:Gammasource}); and the modified Poisson equation, Eq.~(\ref{eq:poissoneq}).
See Appendix~\ref{sec:ppfcode} for more details on the integration.
Note that in~\textsection\ref{sec:PPFST}, we reparametrize $\Gamma$ in terms of a multiplicative correction to the standard Poisson equation $\Sigma(a,k)$.
In the limit of $g=f_{\zeta}=f_G=0$ and, correspondingly, $\Sigma=1$, the PPF perturbation equations recover the $\Lambda$CDM relations.

Finally, the metric ratio $g(a,k)$ can be parametrized by noting that $g(a,k_H=0)\equiv g_{\rm SH}(a)$ and $g(a,k_H=\infty)\equiv g_{\rm QS}(a)$ are scale-independent in the two limiting cases as~\cite{PPF:07}
\bq
 g(a,k) = \frac{g_{\rm SH}(a) + (c_g k_H)^{n_g} g_{\rm QS}(a)}{1 + (c_g k_H)^{n_g}}, \label{eq:gakPPF}
\eq
where $c_g(a)$ and $n_g$ are the interpolation parameters that can be adjusted for each modified gravity theory.
For more details, we refer to~\cite{PPF:07} and Appendix~\ref{sec:ppfcode}.


\subsection{PPF for scalar-tensor theory}\label{sec:PPFST}

PPF fits for minimally and nonminimally coupled scalar-tensor theories or quintessence and generalized/extended quintessence models, respectively, with horizon-scale consistency within the framework of~\cite{PPF:07} have not been developed previously (see, however,~\cite{baker:12} for a description of scalar-tensor models in a different PPF approach).
We shall, therefore, devote this subsection to introduce a calibration method for obtaining consistent near-horizon perturbations within general scalar-tensor models of the form of Eq.~(\ref{eq:action}).
We then discuss the performance of this approach in comparison with the full model-specific fluctuations of the particular scalar-tensor theories.

For our calibration method, we shall use the model-specific linear perturbation theory for scalar-tensor gravity and quintessence models, which we review in Appendix~\ref{sec:STpert}.
Here, we discuss the corresponding quasistatic subhorizon limit thereof, which we use to determine $\gqs(a)$ in Eq.~(\ref{eq:gakPPF}) and the scale dependence of the metric ratio $g(a,k)$.


\subsubsection{The quasistatic subhorizon limit for linear perturbations in scalar-tensor gravity}\label{sec:STQS}

In the quasistatic approximation at subhorizon scales, the perturbed modified Einstein equations of nonminimally coupled scalar-tensor models in their Brans-Dicke representation (see Appendix~\ref{sec:BDpert}) can be simplified to obtain the relations (see, e.g., \cite{espositofarese:00,tsujikawa:08})
\bqa
 g = \frac{\phip}{\phim} & = & -\frac{1}{2\omega+3} \frac{k^2\scal}{a^2 M^2 + k^2 \scal}, \label{eq:gakQSST}\\
 k^2 \phim & = & \frac{\kappa^2}{2 \scal} a^2 \rhm \dm. \label{eq:SigmaQSST}
\eqa
In the limit $k \gg a \, M \, \scal^{-1/2}$, Eq.~(\ref{eq:gakQSST}) becomes scale-independent with the metric ratio $g \rightarrow -1/(2\omega+3)$.
The mass of the scalar field $M$, governing the scale dependence in Eq.~(\ref{eq:gakQSST}), is determined from Eq.~(\ref{eq:BDscalarfield}), replacing $R$ with the trace of Eq.~(\ref{eq:BDEinstein}),
\bq
(2\omega+3) \Box \scal = \kappa^2 T - \omega_{\scal} (\partial_{\alpha}\scal)^2 - 4U + 2 \scal \, U_{\scal}. \label{eq:mass}
\eq
By defining the effective potential $\Box \scal = V_{{\rm eff},\scal}$, the mass of the scalar field is
\bq
 M^2 = \frac{\rmd^2}{\rmd\scal^2} V_{\rm eff}. 
\eq

Together with the conservation equations, Eq.~(\ref{eq:STcons1}), which becomes $\dm' = -k_H \vm$ $(\zeta'=0)$, and Eq.~(\ref{eq:STcons2}), Eqs.~(\ref{eq:gakQSST}) and (\ref{eq:SigmaQSST}) with initial condition $\Delta_{\rm m, i}$ fully define the quasistatic subhorizon linear perturbations, i.e.,
\bqa
 \dm'' + \left( 2 + \frac{H'}{H} \right) \dm' & &  \nonumber \\
 - \frac{3}{2 \scal} \frac{(2\omega+3)a^2 M^2 + (2\omega+4) k^2 \scal}{(2\omega+3)(a^2 M^2 + k^2 \scal)} \frac{H_0^2 \Om}{H^2 a^3} \dm  & = &  0. \nonumber \\
 & &
\eqa


\subsubsection{PPF horizon calibration} \label{sec:STPPF}

Based on the quasistatic subhorizon nonminimally coupled scalar-tensor perturbation limit for the metric ratio, Eq.~(\ref{eq:gakQSST}), and the PPF interpolation formula, Eq.~(\ref{eq:gakPPF}), we extrapolate a PPF fitting function for $g$ for general scalar-tensor theories of the kind considered in Eq.~(\ref{eq:action}) in Brans-Dicke representation.
The quasistatic subhorizon part of our scalar-tensor PPF fit for the metric ratio is obtained from Eq.~(\ref{eq:gakQSST}), hence, we define
\bqa
 \gqs & \equiv &-\frac{1}{2\omega+3}, \\
 c_g & \equiv & \sqrt{\scal} \frac{H}{M}, \\
 n_g & \equiv & 2,
\eqa
in Eq.~(\ref{eq:gakPPF}), which yields the scalar-tensor PPF fit for the metric ratio
\bq
 g(a,k) = \frac{1}{2\omega+3} \frac{a^2 M^2 (2\omega+3) \gsh(a) - \scal \, k^2}{a^2 M^2 + \scal \, k^2}, \label{eq:gakinterpol} \\
\eq
where the mass of the scalar field $M$ is obtained from Eq.~(\ref{eq:mass}).
We determine $\gsh(a)$ by solving the full model-specific linear perturbation theory defined in Appendix~\ref{sec:BDpert} for each nonminimally coupled scalar-tensor model at a given $k_{\rm i}$, which can be chosen through the required range of applicability of the fit.
In this paper, we shall work with $k_{\rm i}=H_0$.
The resulting metric perturbations can be used to obtain the metric ratio $g_{\rm i}(a) = g(a,k_{\rm i})$, which then yields
\bq
 \gsh(a) = g_{\rm i}(a) + \left[ g_{\rm i}(a) + \frac{1}{2\omega+3} \right] \scal \left(\frac{k_{\rm i}}{a M}\right)^2.
\eq

In order to relate the metric to the matter perturbations, we reparametrize the modification to the Poisson equation of the lensing potential in Eq.~(\ref{eq:poissoneq}) by~\cite{amendola:07}
\bq
 k^2 \phim = \Sigma(a,k) \frac{\kappa^2}{2} a^2 \rhm \dm, \label{eq:Sigmapoisson}
\eq
where we have defined
\bq
 \Sigma(a,k)\equiv\frac{1}{1+\Gamma/\phim},
\eq
which reduces to
\bq
 \Sigma(a,k)\simeq\frac{1}{1+f_G}\equiv\Sigma_{\rm QS}
\eq
in the quasistatic subhorizon limit, recovering Eq.~(\ref{eq:qspoissoneq}).
This sets the PPF parameter
\bq
 f_G = \scal - 1
\eq
from Eq.~(\ref{eq:SigmaQSST}) for nonminimally coupled scalar-tensor models, i.e., $\Sigma_{\rm QS} = \scal^{-1}$.
In order to obtain $\Sigma(a,k)$ on super- and near-horizon scales, we use the full model-specific perturbations at $k_{\rm i}$ used in the determination of $\gsh(a)$ in Eqs.~(\ref{eq:Gamma}) and (\ref{eq:Gammasource}) to obtain $f_{\zeta}(a)$.
Note that we refrain here from using simple numbers or scalings for $f_{\zeta}(a)$ based on $\gsh(a)$, providing a simple and general approach for determining $f_{\zeta}(a)$.

The PPF function $f_{\zeta}$ determined by this procedure together with $f_G$, $g$, and $w_{\rm eff}$, which is obtained from the background equations, Eqs.~(\ref{eq:BDfriedmann1}) to (\ref{eq:BDenmomcons}), fully determine the linear PPF perturbations corresponding to the model-specific perturbations of the nonminimally coupled scalar-tensor theory of Appendix~\ref{sec:BDpert} up to the transition scale $c_{\Gamma}$, which remains the only free PPF parameter that has to be calibrated to obtain the correct scale dependence in the modification of the Poisson equation of the lensing potential, Eq.~(\ref{eq:Sigmapoisson}).

For minimally coupled models during matter domination, the metric ratio is
\bq
 g=0,
\eq
which follows from $F={\rm const}$ and the Einstein equation $\phip = -\delta F/F=0$.
From this and Eq.~(\ref{eq:mincouppoisson}), it follows that
\bq
 f_G=0
\eq
in minimally coupled models.
However, in general, $\Sigma\neq1$ at near-horizon scales since dark energy may cluster at the largest scales.
Note that for minimally coupled models, the Poisson equation should not be regarded as modified in the sense of changing gravity but contributions of scalar-field perturbations are parametrized as an effective modification to the relation between metric and matter perturbations. 
By calibrating $\cG$, we can apply the same procedure as described here for the nonminimally coupled scalar-tensor theories to obtain $f_{\zeta}(a)$ at $k_{\rm i}$ for the quintessence models and, hence, determine the linear PPF perturbations corresponding to the full quintessence perturbations of Appendix~\ref{sec:QSpert}.


\subsubsection{Performance of PPF fits} \label{sec:PPFperformance}

\begin{figure*}
 \resizebox{0.95\hsize}{!}{\includegraphics{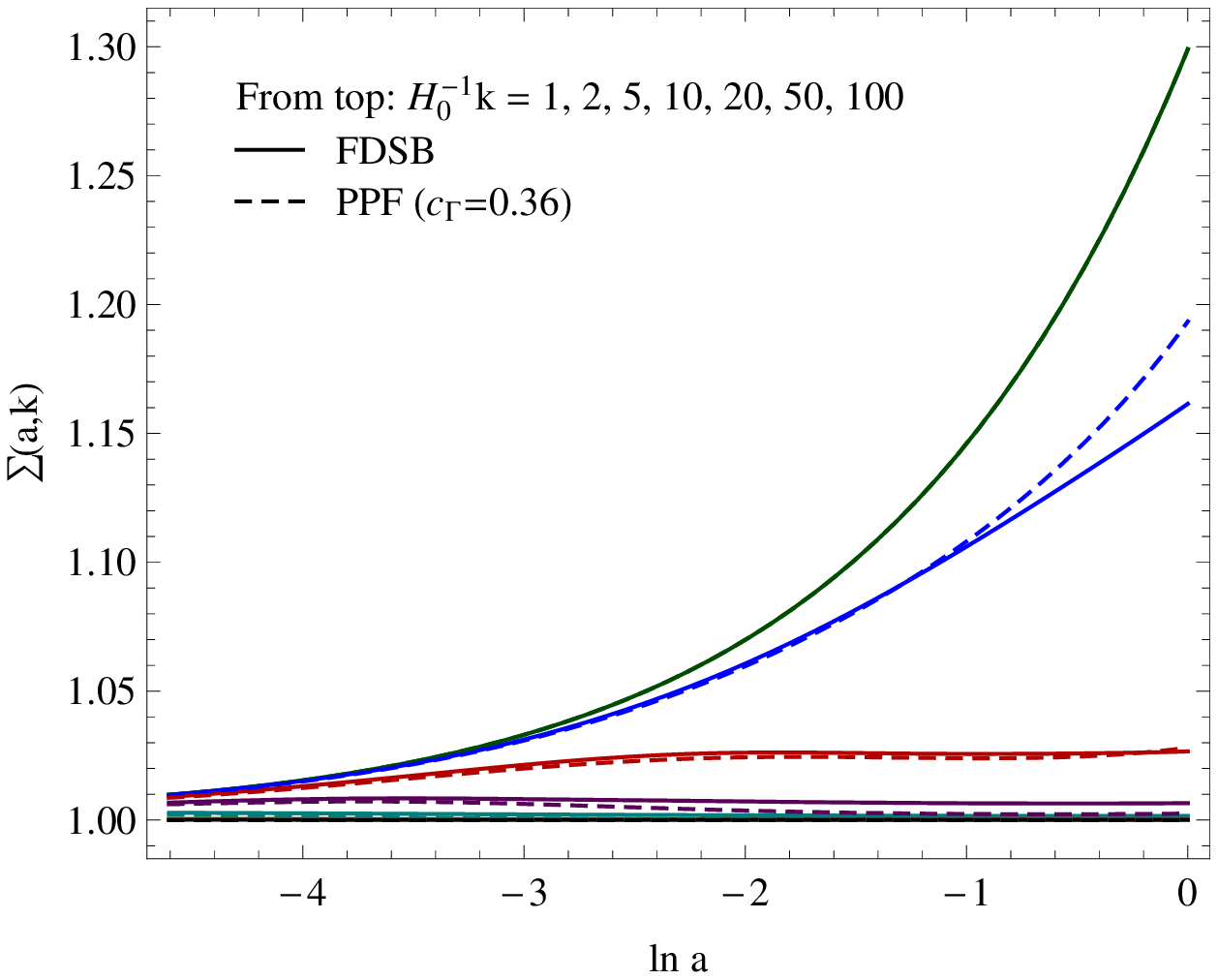}\includegraphics{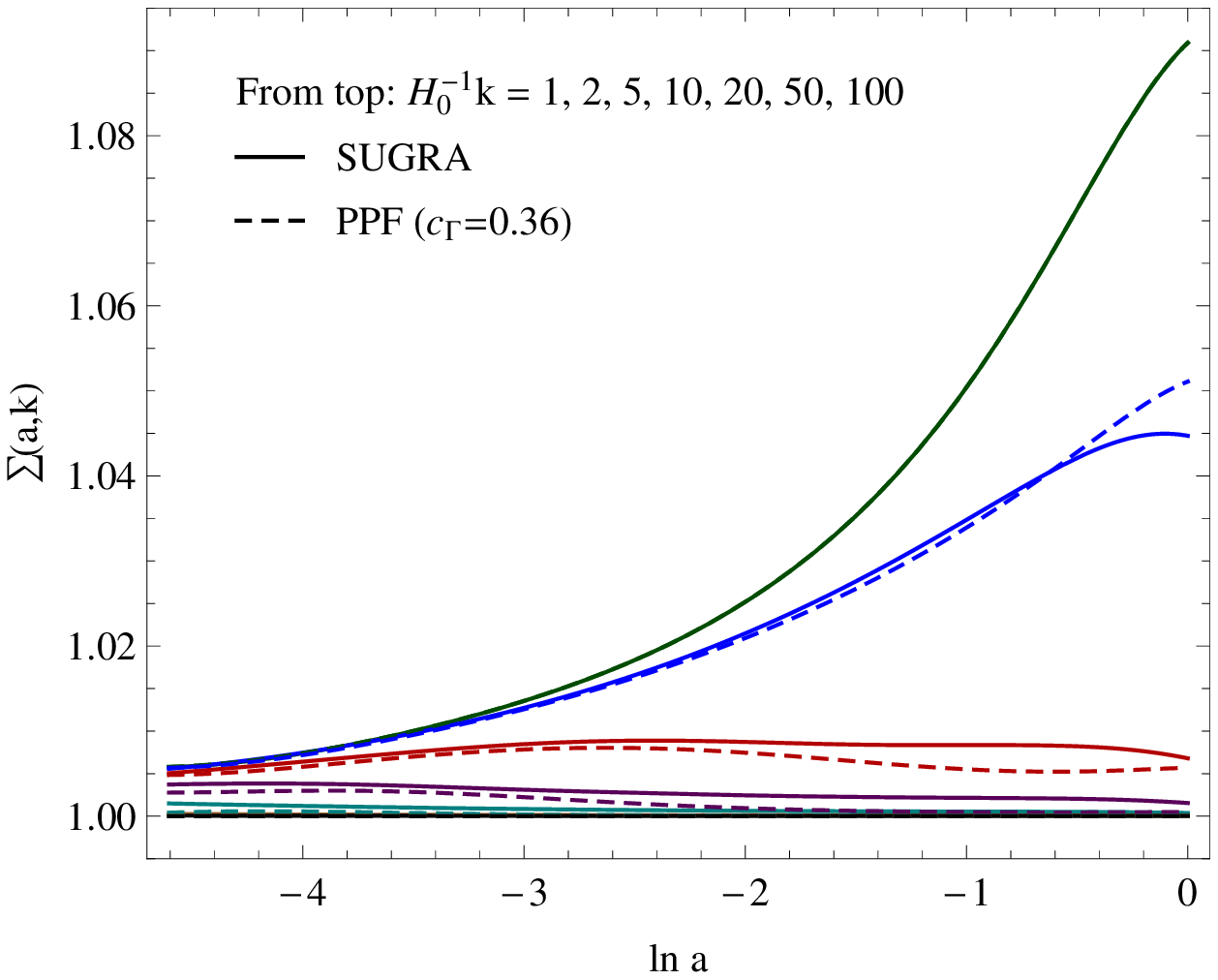}}
 \resizebox{0.95\hsize}{!}{\includegraphics{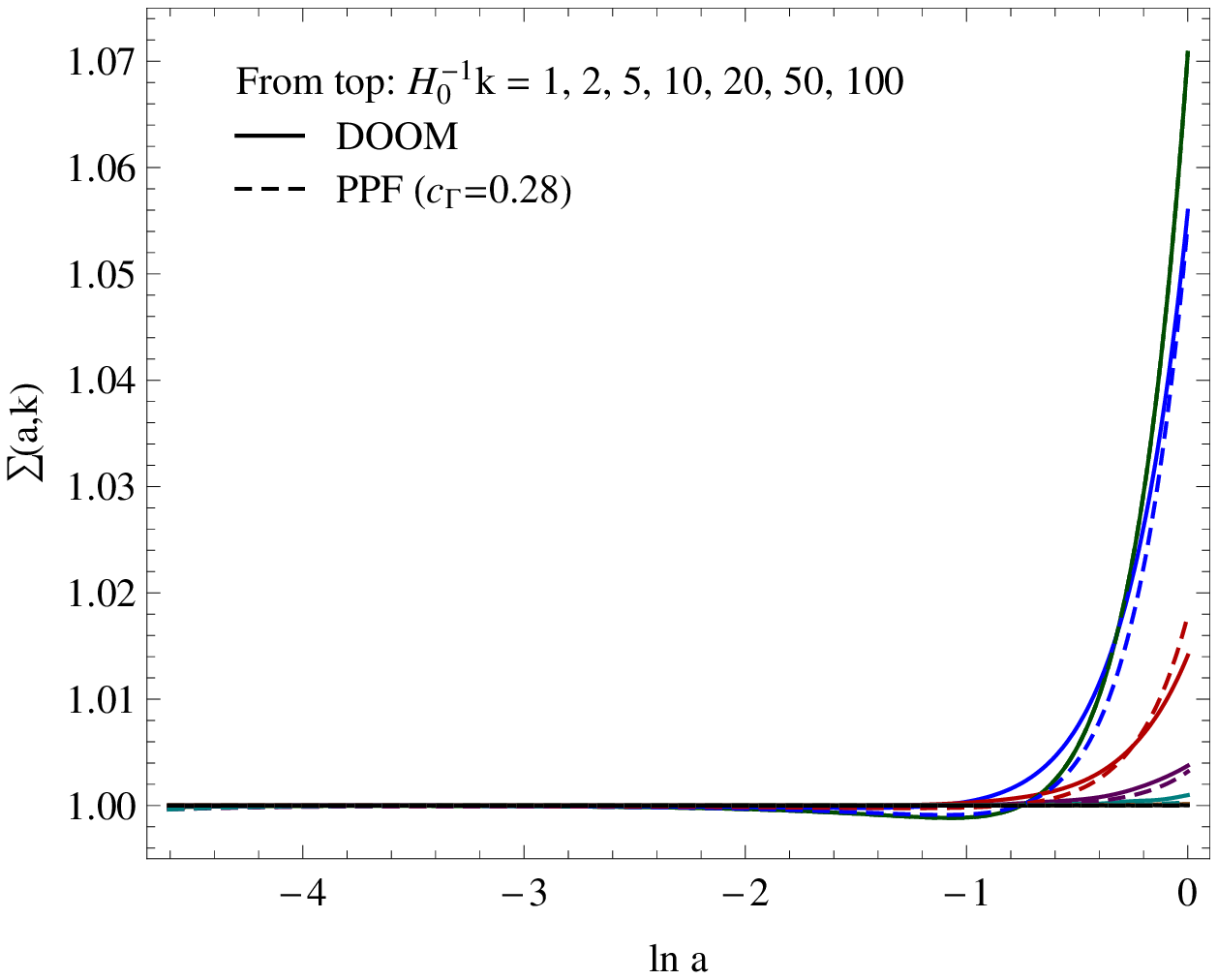}\includegraphics{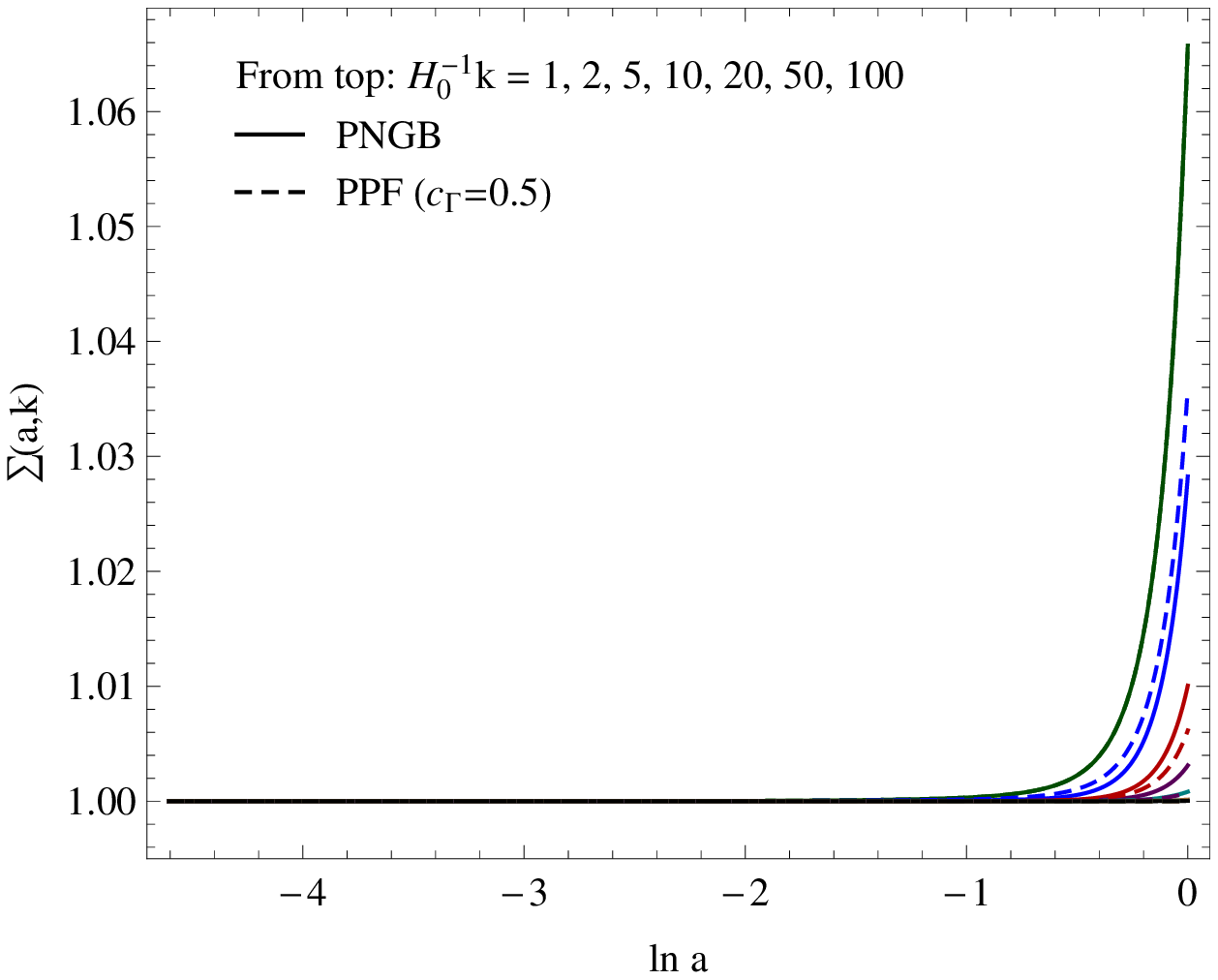}}
 \resizebox{0.95\hsize}{!}{\includegraphics{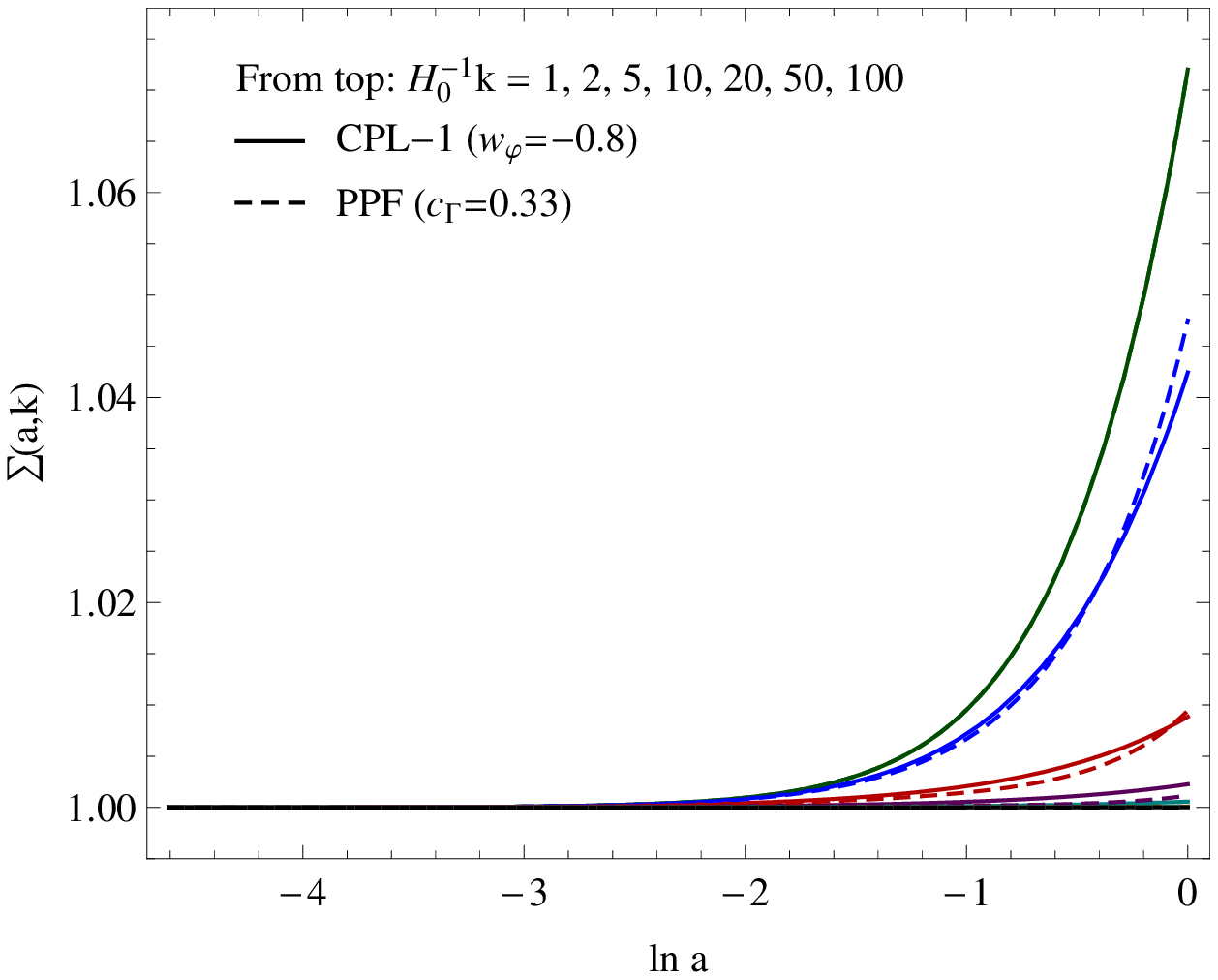}\includegraphics{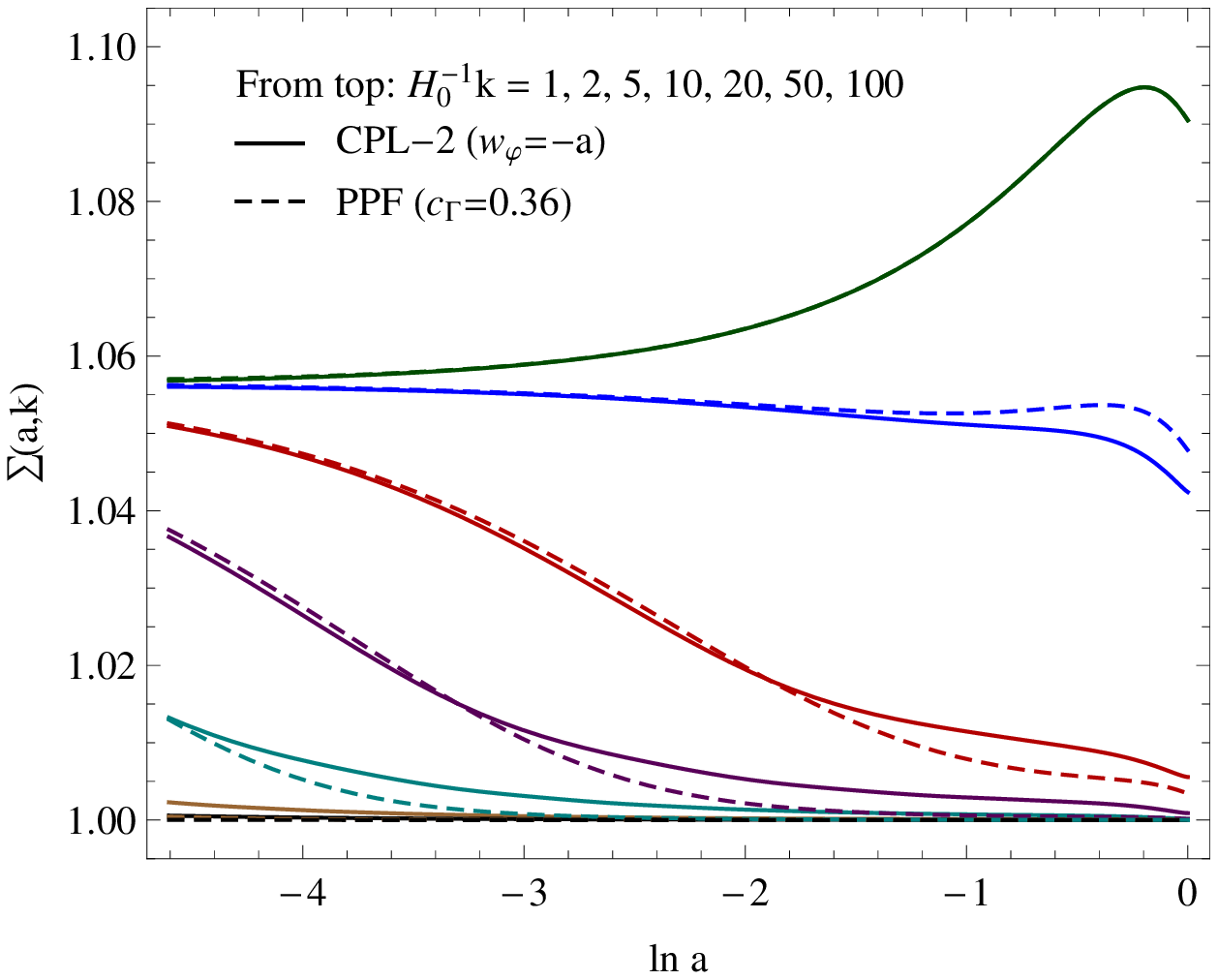}}
\caption{Modification $\Sigma$ of the Poisson equation, Eq.~(\ref{eq:Sigmapoisson}), relating the lensing potential to the matter density fluctuation, due to contributions of the scalar field ($\Sigma=1$ in $\Lambda$CDM). We show the clustering of the scalar field for the six minimally coupled scalar-tensor (or quintessence) models defined in~\textsection\ref{sec:quintessence} at large scales. The quintessence field does not introduce anisotropic stress ($g=0$). Given the background evolution, the two PPF functions completely define the quintessence perturbations. The PPF fits developed in~\textsection\ref{sec:STPPF} (dashed lines) provide good fits to the model-specific quintessence perturbations (solid lines). The transition from the quasistatic subhorizon to the superhorizon limits, governed by the PPF parameter $\cG$, is calibrated at $k_{\rm i}=H_0$, where the fit overlaps with the predictions from the full perturbations.
Note that $\Sigma$ does not recover the $\Lambda$CDM limit at early times for the CPL-2 model due to the strong background contribution of the scalar field at small $a$.
}
\label{fig:Sigma_quintessence}
\end{figure*}

\begin{figure*}
 \resizebox{\hsize}{!}{\includegraphics{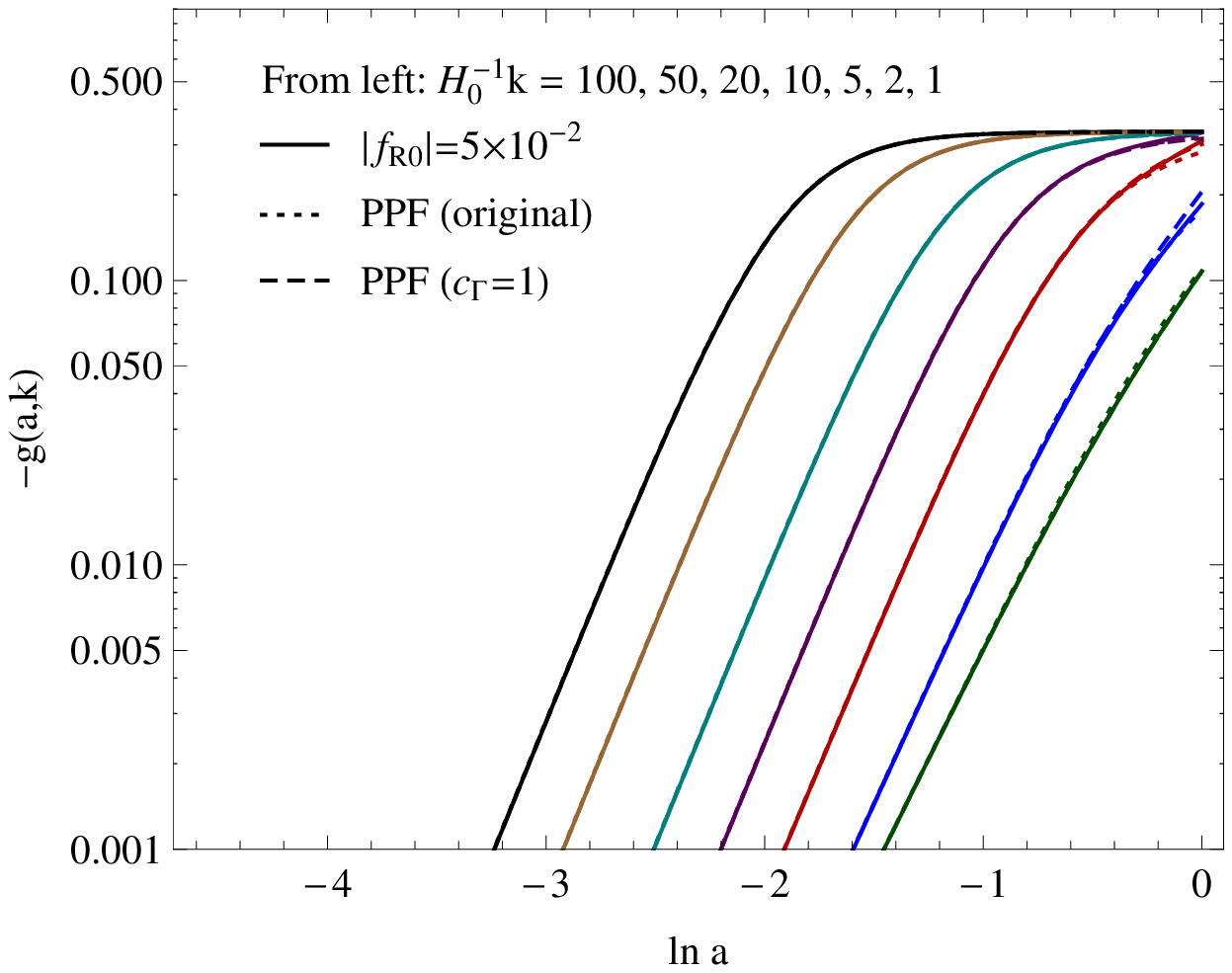}\resizebox{0.687\hsize}{!}{\includegraphics{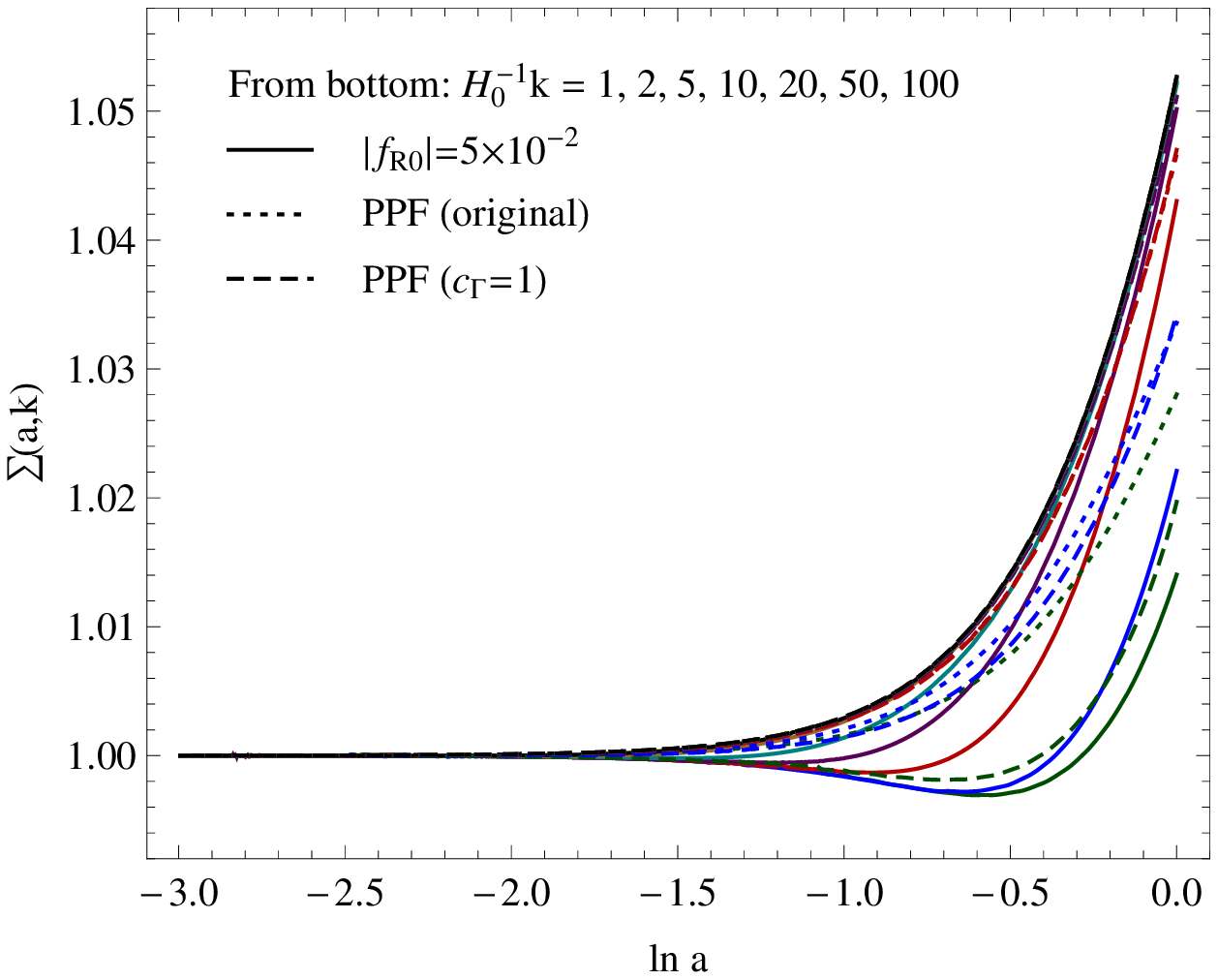}}}
 \resizebox{\hsize}{!}{\includegraphics{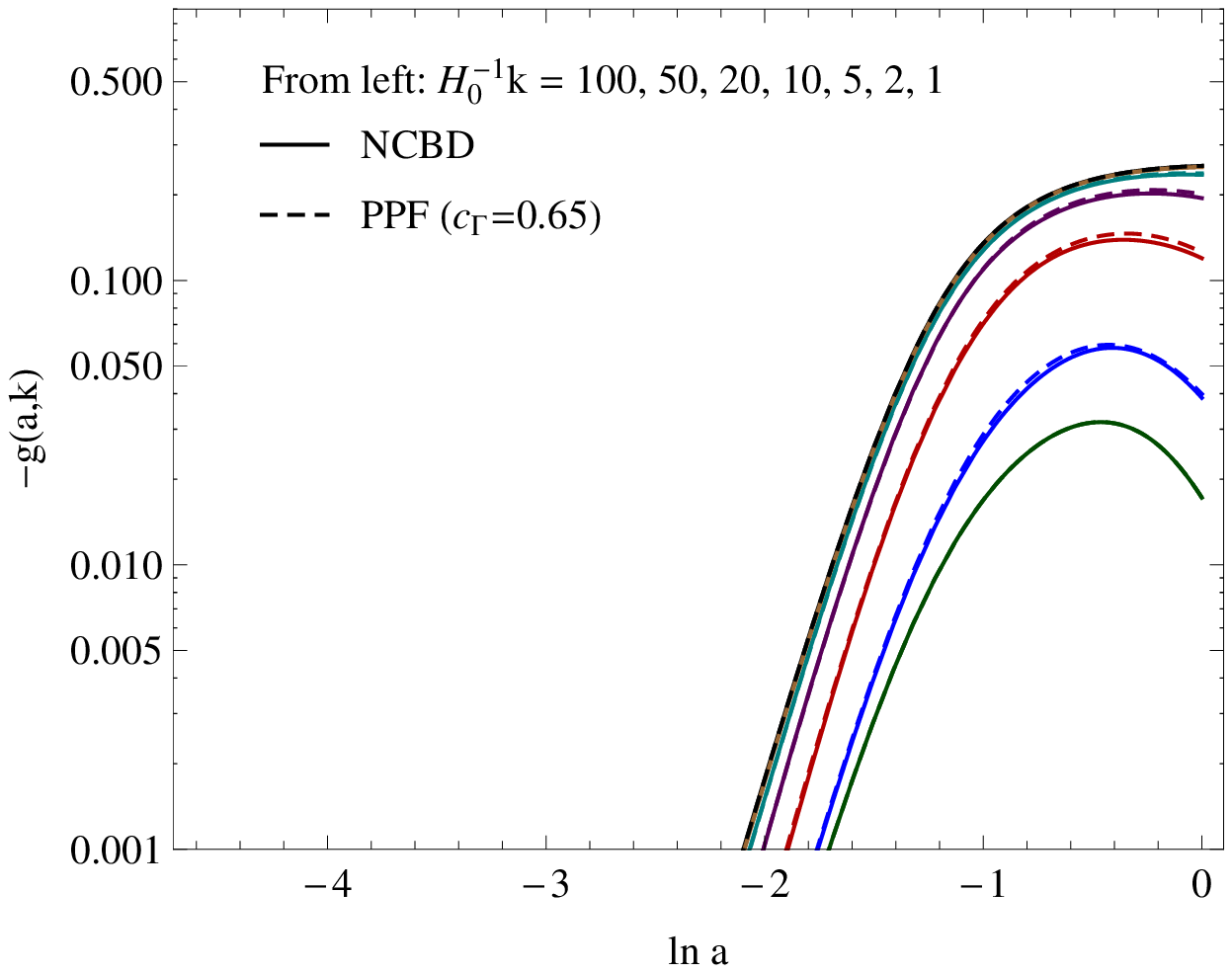}\resizebox{0.687\hsize}{!}{\includegraphics{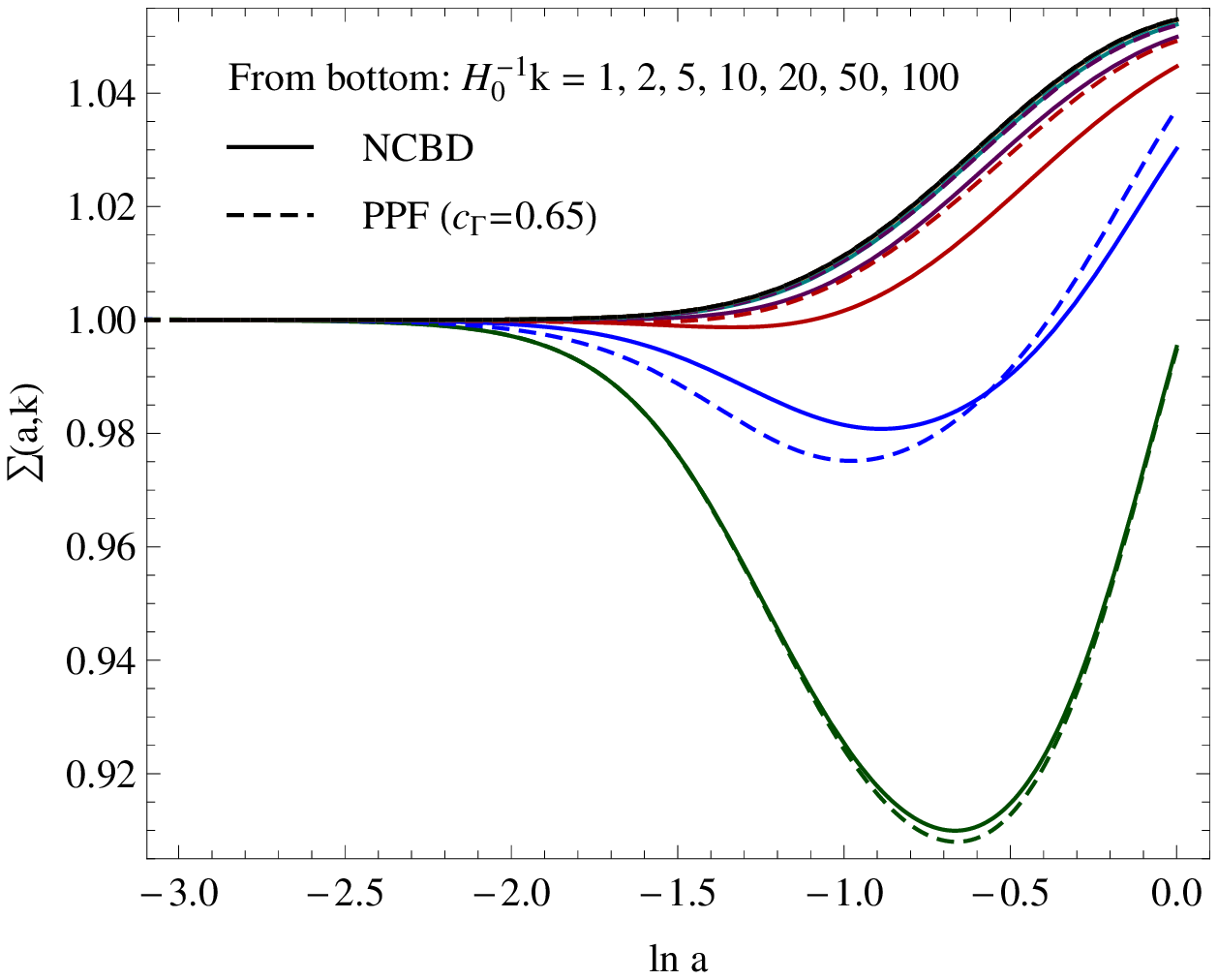}}}
\caption{Metric ratio $g$ describing the effective anisotropic stress (left-hand column) and the modification of the Poisson equation $\Sigma$ (right-hand column) for the designer $f(R)$ gravity (\textsection\ref{sec:fRgravity}) (upper row) and the NCBD model defined in Eqs.~(\ref{eq:NCBD1}) and (\ref{eq:NCBD2}) (\textsection\ref{sec:NCBD}) (lower row).
The calibration technique developed in~\textsection\ref{sec:STPPF} produces a slightly improved fit for $\Sigma$ compared to the original $f(R)$ PPF fit and is governed by the PPF parameter $\cG$. Note that for $f(R)$ gravity with small values of $|f_{R0}|$, the degree of correspondence between the calibrated and original PPF as well as the `true' functions $g$ and $\Sigma$ improve.
}
\label{fig:PPF_NM}
\end{figure*}

We test our PPF fits for the metric ratio $g(a,k)$ and the modification of the Poisson equation $\Sigma(a,k)$ derived in~\textsection\ref{sec:STPPF} against their predictions from the model-specific linear perturbations of the quintessence models described in Appendix~\ref{sec:QSpert} (see Fig.~\ref{fig:Sigma_quintessence}) and the nonminimally coupled scalar-tensor theories of Appendix~\ref{sec:BDpert} (see Fig.~\ref{fig:PPF_NM}).
For the cosmological and model parameters, we adopt the values defined in~\textsection\ref{sec:DEmodels}.
In the case of $f(R)$ gravity, we also give the original PPF fit of~\cite{PPF:07} (see Appendix~\ref{sec:fRPPF}) in the first row of Fig.~\ref{fig:PPF_NM}, which is in good agreement with the results from~\textsection\ref{sec:STPPF}.
Note that, due to our calibration of $f_{\zeta}$ at $k_{\rm i}=H_0$, our fit captures the `true' $\Sigma(a,k)$ slightly better than the original $f(R)$ PPF fits of~\cite{PPF:07}, which use $f_{\zeta} = -\gsh/3$.
We choose a rather large value of $|f_{R0}|$ to highlight differences between the different PPF predictions as well as the full perturbation theory for $f(R)$ gravity.
The deviations become smaller with smaller values of $|f_{R0}|$.
Note, however, that an upper bound of $|f_{R0}|\lesssim0.35$ and $|f_{R0}|\lesssim0.07$ (95\% confidence level) is inferred from CMB only and in combination with galaxy-ISW data~\cite{lombriser:10}, respectively.
For $|f_{R0}|=0.05$ chosen in Fig.~\ref{fig:PPF_NM}, at the horizon today, the metric ratio can be of order $g\simeq0.1$, which is a factor of $\sim2$ larger than the result we would obtain if ignoring the superhorizon contribution by setting $\gsh=0$ in Eq.~(\ref{eq:gakPPF}).

Our PPF fits are in good agreement with the model-specific perturbations of the minimally and nonminimally coupled scalar-tensor theories.
Note that there remains potential to improve the correspondence in $\Sigma(a,k)$ by introducing time and/or scale dependence in the transition scale $\cG$.


\subsubsection{Correspondence of PPF for $f(R)$ gravity}\label{sec:fRlimit}

$f(R)$ gravity is equivalent to a nonminimally coupled scalar-tensor theory with the specifications given in Eqs.~(\ref{eq:fRreplace1}), (\ref{eq:fRreplace2}), and (\ref{eq:fRreplace3}).
Hence, the PPF fits for general nonminimally coupled scalar-tensor theories constructed in~\textsection\ref{sec:PPFST} should reproduce the $f(R)$ PPF description in the corresponding limit.

In $f(R)$ gravity, the scalar-field equation is obtained from setting $\omega=0$, $\scal=1+f_R$, and $U=(R \, f_R - f)/2$ in Eq.~(\ref{eq:mass}), hence,
\bq
 \Box f_R = \frac{1}{3} \left[ (1-f_R) R + 2f + \kappa^2 T \right] \equiv \frac{\partial V_{\rm eff}}{\partial f_R}
\eq
which defines $f_G=f_R$ and the mass of the $f_R$ field
\bq
 M^2_{f_R} = \frac{1}{3} \left( \frac{1+f_R}{f_{RR}} - R \right).
\eq
At high curvatures,
\bq
 M^2_{f_R} \simeq \frac{1}{3}\frac{1+f_R}{f_{RR}} = \frac{1}{3} \frac{H \, R'}{B \, H'},
\eq
where for a background evolution recovering a $\Lambda$CDM expansion history, this becomes $M^2_{f_R}=2H^2/B$ with $B$ determined by Eq.~(\ref{eq:compton}).
For the interpolation weights, this yields
\bq
 c_g = (1+f_R) \sqrt{\frac{B}{2}} \simeq 0.71\sqrt{B} \\
\eq
for $|f_R| \ll 1$ with $n_g=2$.
Since $\omega=0$, we have $g_{\rm QS}=-1/3$ in Eq.~(\ref{eq:gakQSST}).
This recovers the PPF fits of~\cite{PPF:07} at small and intermediate scales (see Appendix~\ref{sec:fRPPF}).

We make comparisons of the super-, near-, and subhorizon PPF fits in~\textsection\ref{sec:PPFperformance}.
Note that at early times, we can neglect the time derivative in Eq.~(\ref{eq:fRSHrel}), which leads to $\gsh \rightarrow -B/2$.
Using this approximation, we may obtain a simplified superhorizon PPF fit for $g$ without solving Eq.~(\ref{eq:fRSH}) through
$\gsh \simeq -(B/2) \Om(a)^n$ with $\Om(a) \equiv (H_0/H)^2\Om a^{-3}$, where $n \in (0.5,0.6)$ is found here to give good agreement with the full perturbations.
Here, we shall, however, not use this approach but we remind the reader that simple approximations of this form can be very useful in parameter estimation analyses employing PPF Boltzmann linear theory solvers~\cite{fang:08a,fang:08b,daniel:10,hojjati:11,dossett:11b} in a Markov chain Monte-Carlo exploration of the cosmological parameter space to obtain constraints on the gravitational models.
For this purpose, the perturbed field equations need to be solved several thousand times.
The PPF approach can significantly increase the efficiency of the integration.
For an application of the PPF framework in $f(R)$ gravity model constraints, see, e.g.,~\cite{lombriser:10}.


\subsection{PPF as phenomenology} \label{sec:ppftestmod}

In order to study more general effects on galaxy clustering caused by deviations from the concordance model, in addition to the scalar-tensor and DGP models, we further construct two phenomenological modifications of gravity (or nonstandard cosmological models) based on the PPF formalism described in~\textsection\ref{sec:ppf}.
We assume for both models a standard quasistatic subhorizon Poisson equation or negligible modifications thereof, $f_G=0$, and, furthermore, $f_{\zeta}=0$ with $\cG=1$.

In the first model (PHEN-1), we assume a $\Lambda$CDM expansion history, $w_0=-1$ and $w_a=0$ in Eq.~(\ref{eq:rhomod}), and an effective superhorizon anisotropic stress described by $\gsh(a) = g_0 a^3$ such that modifications vanish at early times.
We further choose $\gqs=0$, $c_g=1$, and $n_g=2$ in Eq.~(\ref{eq:gakPPF}) such that the modifications of the model are defined by
\bq
 g(a,k) = \frac{g_0 a^{3}}{1+k_H^2}, \label{eq:PPFpheno1}
\eq
where we set $g_0=10$ for illustration.
Note that in such a model, departures of $\Lambda$CDM disappear on subhorizon scales, i.e., $g\lesssim10^{-3}$ at $k \geq 10^2 H_0$ today, whereas $g=5$ at $k=H_0$.

Our second phenomenological model (PHEN-2) is motivated by the results of~\cite{lombriser:11a}.
The expansion history is defined through the effective dark energy equation of state in CPL form, i.e., $w_0$ and $w_a$ in Eq.~(\ref{eq:rhomod}).
We further define $f_G=f_{\zeta}=0$ and $c_{\Gamma}=1$.
The metric ratio is chosen to be a constant,
\bq
 g(a,k) = g_0, \label{eq:PPFpheno2}
\eq
i.e., modifications are also present at the largest scales and early times.
We motivate our choice of $w_0$, $w_a$, and $g_0$ by the best-fit parameters found in~\cite{lombriser:11a}, i.e., $w_0\simeq-1.15$, $w_a\simeq1.17$, and $g_0\simeq-0.15$.
Remarkably, such a model is consistent with current cosmological observations on linear scales, simultaneously including all measured multipoles of the CMB data, their cross-correlation with foreground galaxies through the ISW effect, as well as geometric, weak lensing, and clustering probes while slightly increasing the best-fit likelihood and enhancing the average likelihood over the posterior distribution.
The corresponding marginalized likelihoods are consistent with $\Lambda$CDM.
The key feature of this strong nonstandard consistency is a degeneracy between the early anisotropic stress and $w_a$ when $w_a \approx -w_0$ in the CMB data.
Note, however, that other cosmological parameters like the matter density $\Omega_{\rm m}$ are not set to the best-fit values of~\cite{lombriser:11a} but are fixed to the same values that are used here for all models (see~\textsection\ref{sec:DEmodels}) and, hence, deviations from $\Lambda$CDM predictions may appear stronger.

We stress that the phenomenological models given here may not be physical, and it is not clear whether more rigorous theoretical models with these properties may be constructed.
The PPF formalism, however, offers the possibility of testing such nonstandard cosmologies while retaining energy-momentum conservation of the fluid contributions and avoiding gauge artifacts as apparent deviations from $\Lambda$CDM.
Despite the possibility of these phenomenological modifications to be unphysical, if $\Lambda$CDM is the correct model, we ultimately aim at ruling out such deviations through observations.
Moreover, such modifications offer a great opportunity to test the constraining power of future surveys.


\section{Signatures of modified gravity and dark energy models in galaxy clustering} \label{sec:observables}

With the PPF fits for modified gravity and dark energy models developed in~\textsection\ref{sec:ppf}, we consistently compute the galaxy power spectrum from the subhorizon to the near-horizon scales.
In~\textsection\ref{sec:relcorr}, we briefly review the relativistic formula for the observed galaxy fluctuation field and discuss its applications to modified gravity and dark energy models.
In~\textsection\ref{sec:significance}, we estimate the detection significance of those models expected with future galaxy surveys.


\subsection{Relativistic effects in galaxy clustering} \label{sec:relcorr}

\begin{figure*}
 \resizebox{0.495\hsize}{!}{\includegraphics{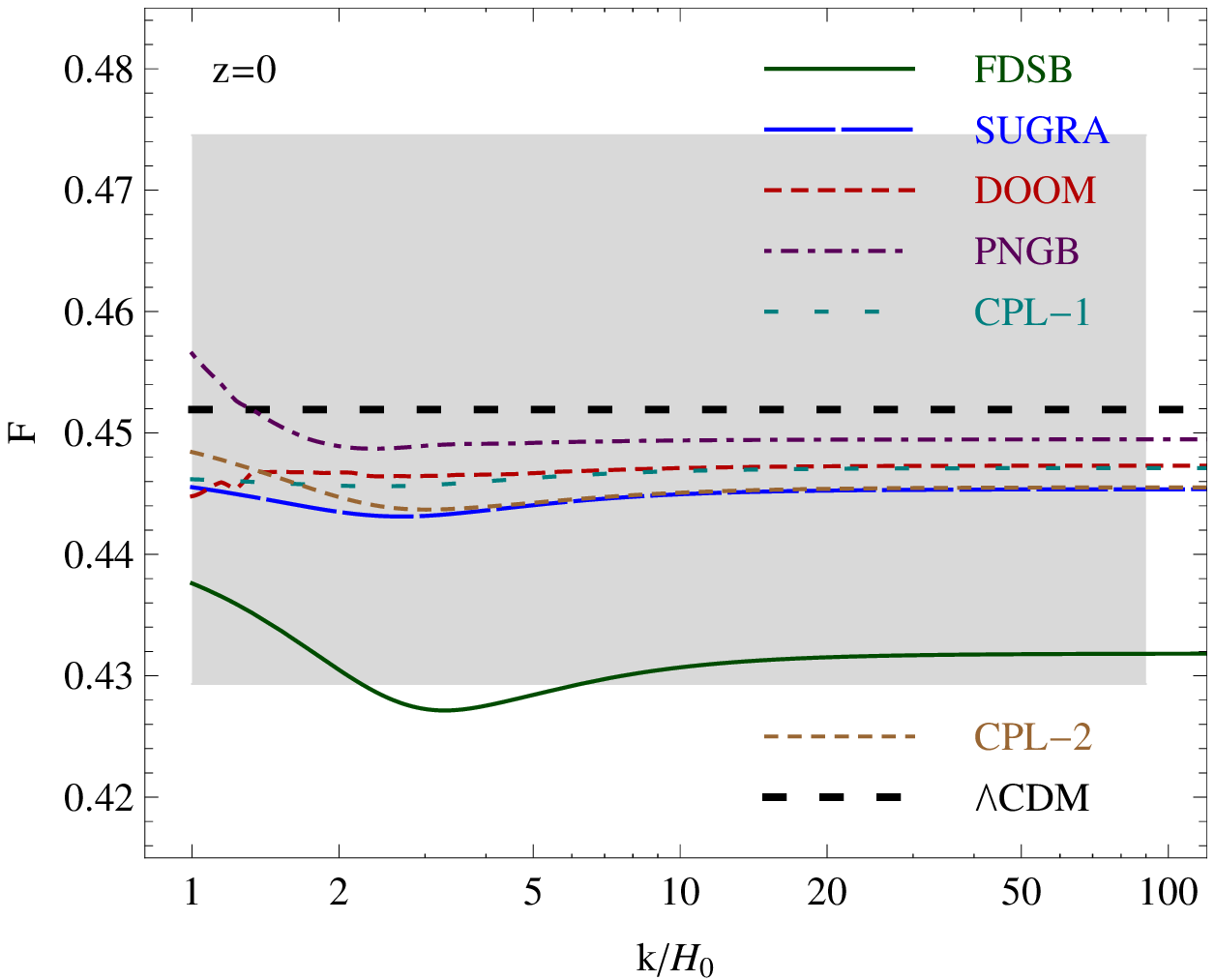}}\resizebox{0.5\hsize}{!}{\includegraphics{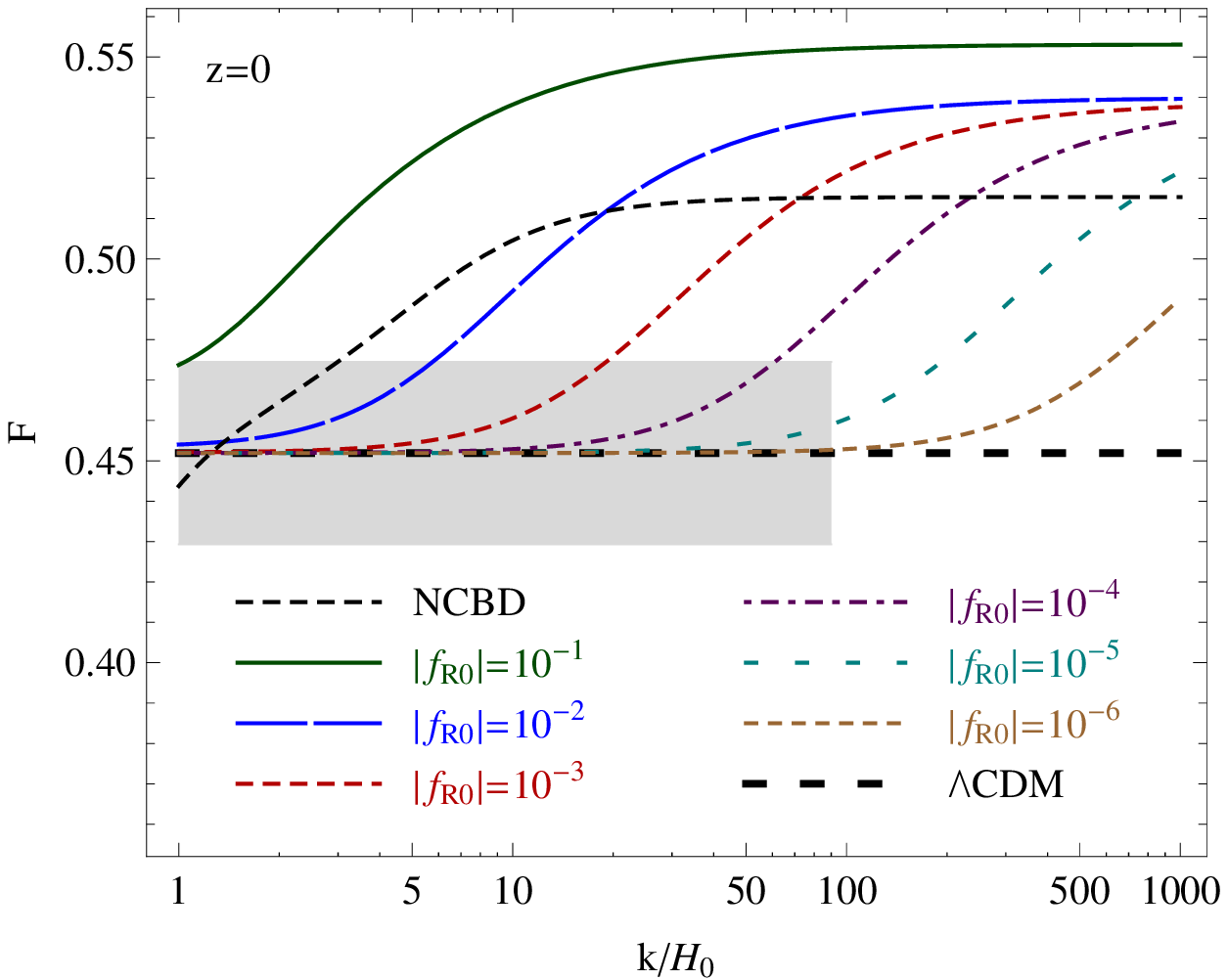}}
\caption{Newtonian growth rate of structure $\mathcal{F}$ at $z=0$ in Eq.~(\ref{eq:newtf}) for the minimally coupled (left-hand panel) and nonminimally coupled (right-hand panel) scalar-tensor theories in~\textsection\ref{sec:st}. 
When the comoving-gauge curvature is conserved $(\zeta'=0)$, the Newtonian growth rate~$\mathcal{F}$ reduces to the usual logarithmic growth rate~$f$ of structure in Eq.~(\ref{eq:logf}).
The gray shaded region indicates the prediction of 1-$\sigma$ measurement uncertainties from a multitracer analysis of future galaxy surveys (see~\textsection\ref{sec:significance}).
}
\label{fig:FST}
\end{figure*}

\begin{figure}
 \resizebox{\hsize}{!}{\includegraphics{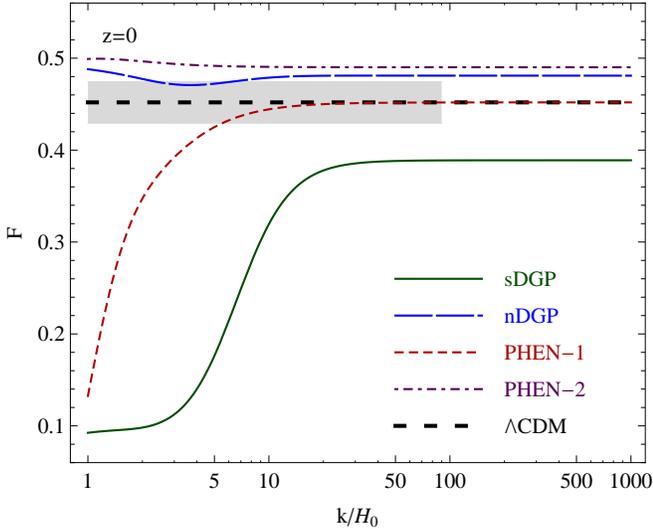}}
\caption{Same as Fig.~\ref{fig:FST} but for the DGP and phenomenological models described in~\textsection\ref{sec:DGP} and~\textsection\ref{sec:pheno}, respectively.}
\label{fig:Fphen}
\end{figure}

\begin{figure*}
 \resizebox{0.9\hsize}{!}{\includegraphics{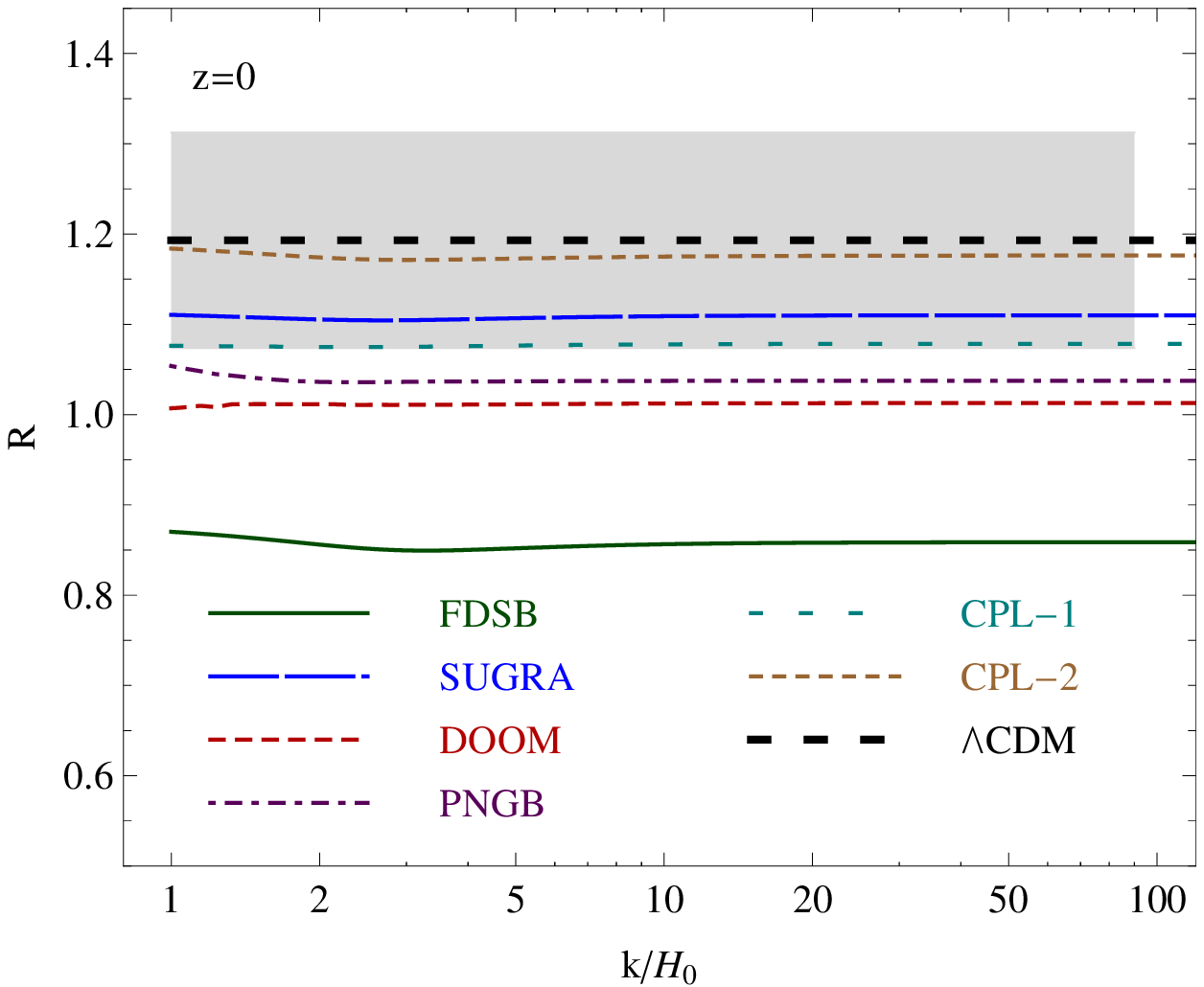}\includegraphics{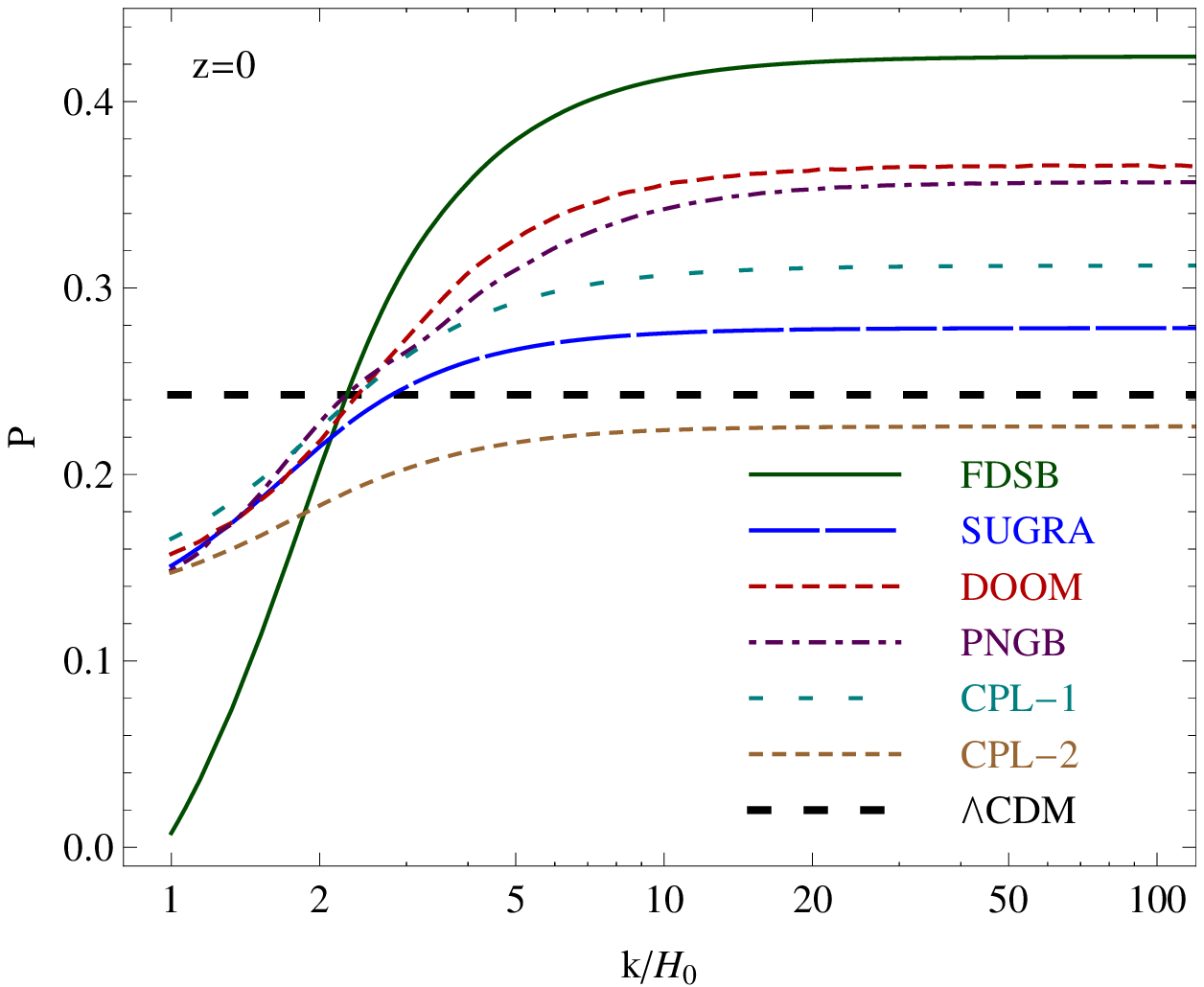}}
 \resizebox{0.45\hsize}{!}{\includegraphics{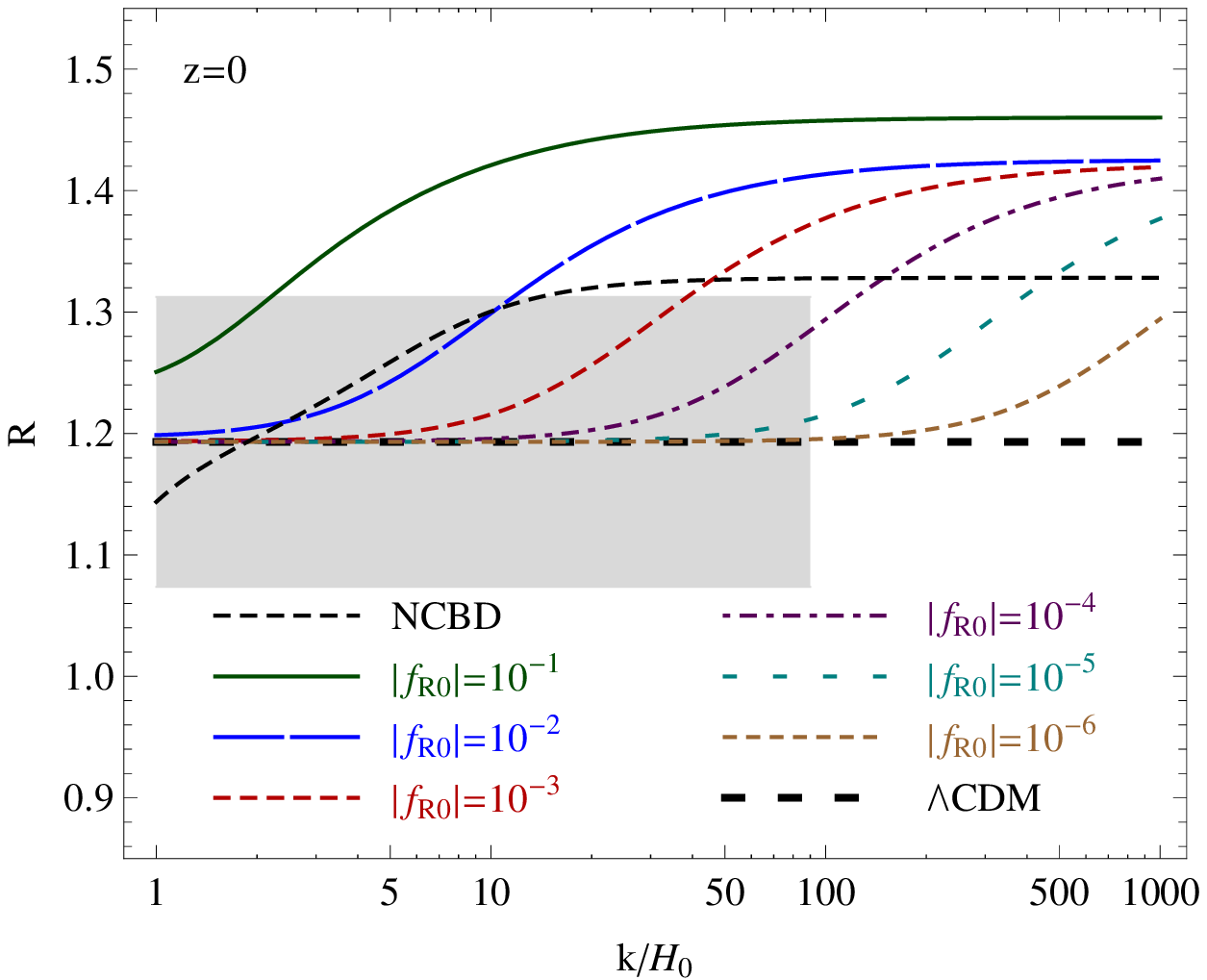}}\resizebox{0.46\hsize}{!}{\includegraphics{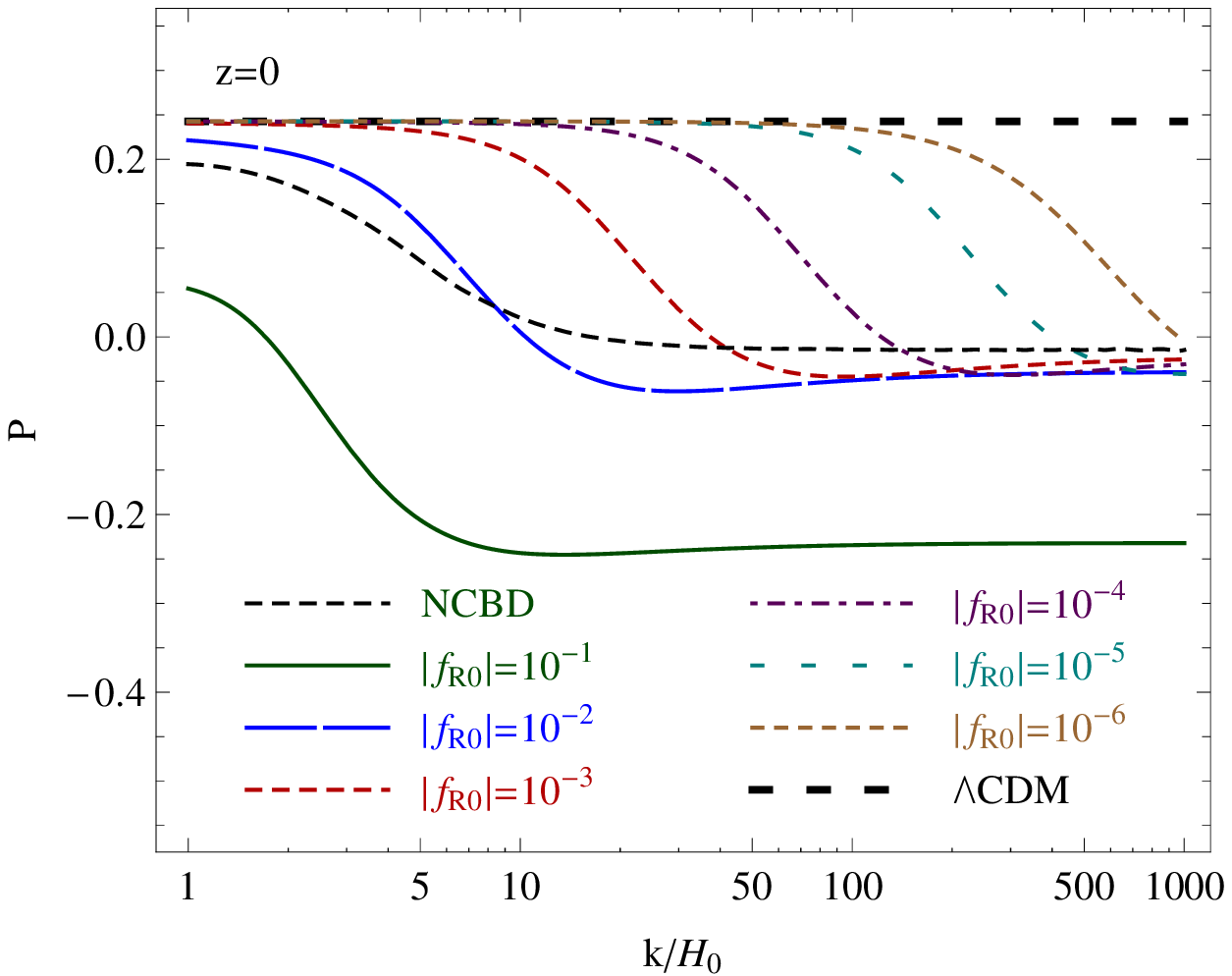}}
 \resizebox{0.9\hsize}{!}{\includegraphics{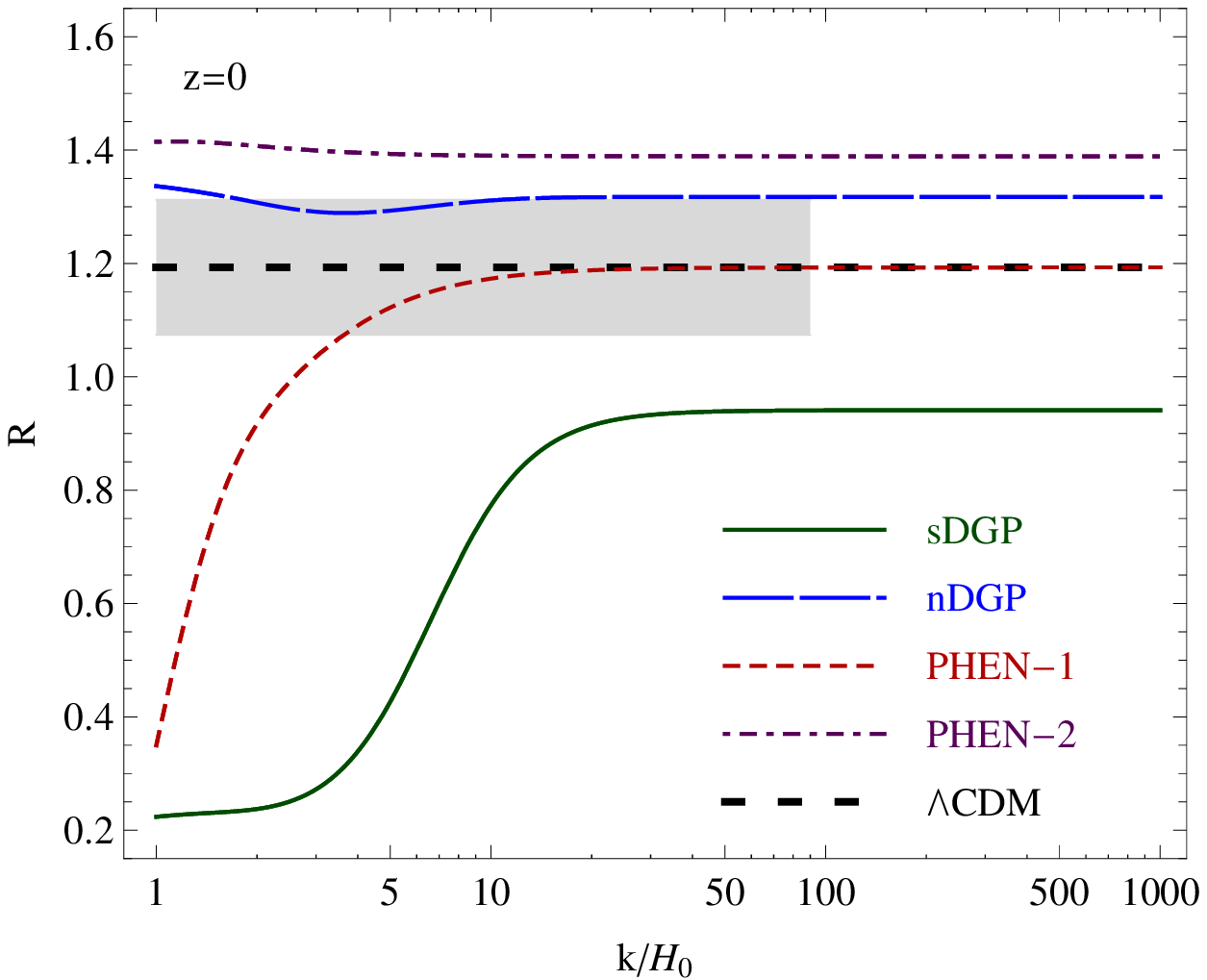}\includegraphics{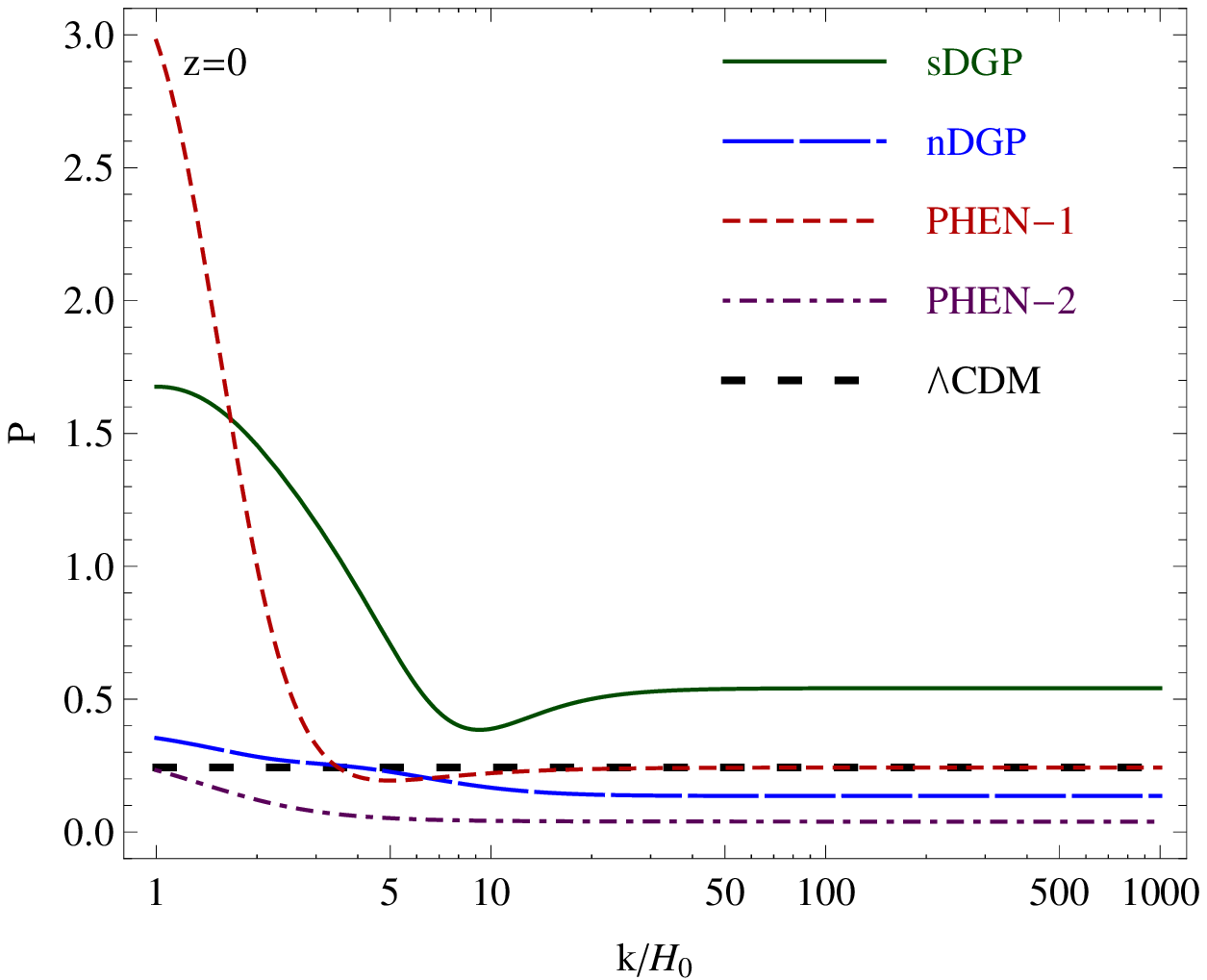}}
\caption{Relativistic contributions to galaxy clustering for
the modified gravity theories and dark energy models at $z=0$. Compared to the Newtonian description, there exist two additional terms $\mathcal{R}$ (left-hand column) and $\mathcal{P}$ (right-hand column) that characterize the relativistic effects described in Eq.~(\ref{eq:reldg}).
From the top row to the bottom row, each panel shows $\mathcal{R}$ and $\mathcal{P}$ for the minimally coupled (top row) and the nonminimally coupled (middle row) scalar-tensor models in~\textsection\ref{sec:st} and the DGP~(\textsection\ref{sec:DGP}) and phenomenological~(\textsection\ref{sec:pheno}) models (bottom row).
Since $\mathcal{R}\propto\mathcal{F}$ in Eq.~(\ref{eq:RFprop}), the overall shape of~$\mathcal{R}$ is similar to that of~$\mathcal{F}$ in Figs.~\ref{fig:FST} and~\ref{fig:Fphen}, but slightly differs due to the difference in the background evolution.
The gray-shaded region indicates the prediction of 1-$\sigma$ measurement uncertainties from a multitracer analysis of future galaxy surveys (see~\textsection\ref{sec:significance}). 
Since $\sigma_{\mathcal{P}}/\mathcal{P}\simeq10$, we do not show the 1-$\sigma$ deviation for $\mathcal{P}$. 
}
\label{fig:RPphen}
\end{figure*}

The relativistic description of galaxy clustering has been developed in the past few years~\cite{yoo:09,yoo:10,bonvin:11,challinor:11,bruni:11,baldauf:11,jeong:11,yoo:12}.
Given the observed redshift and the observed galaxy position on the sky, the full relativistic formula for the observed galaxy fluctuation can be derived by tracing the photon path backward in time.
Since the observed quantities and the photon path are affected by not only the matter fluctuations but also the relativistic contributions of the gravitational potential, the relations among the source, the observer, and the metric perturbations are nontrivial and they require a relativistic treatment for solving the geodesic equation. 
In this manner, the full relativistic description provides a complete picture of all the effects in galaxy clustering such as the redshift-space distortion, the gravitational lensing, the Sachs-Wolfe effect, and their relativistic corrections. Furthermore, this procedure is necessary because all these effects are not individually separable, and what we measure in galaxy surveys are the number of galaxies, their observed redshifts, and the positions on the sky, not the individual effects.
We refer to~\cite{yoo:09,yoo:10} for the derivation and subtle gauge issues associated with it.

The matter and metric perturbations along the photon path affect the observed galaxy fluctuation.
Further contributions arise from the vector and the tensor perturbations.
In generic modified gravity theories, the vector perturbations decay with time as in the standard $\Lambda$CDM model.
Furthermore, the primordial tensor perturbations are well constrained by the CMB observations.
Therefore, in modified gravity theories and dark energy models that mimic the $\Lambda$CDM model at early times, we can safely ignore the vector and the tensor contributions.
With only the scalar contributions, the observed galaxy fluctuation can be obtained as \cite{yoo:09,yoo:10,yoo:12} 
\bqa
 \Delta^{\rm obs}_{\rm g} & = & b \, \dm - e \, \delta z_v + \Psi + 2 \Phi + V + 3\delta z_{\chi} \nonumber \\
 & & + 2\frac{\delta r_{\chi}}{r} - 2\mathcal{K} - H \frac{\rmd}{\rmd z} \left( \frac{\delta z_{\chi}}{a \, H} \right) - 5p \, \delta \mathcal{D}_L, \label{eq:obsgalfluct}
\eqa
where $b$ is the linear bias factor, the evolution bias factor $e = \rmd \ln \bar{n}_g/\rmd \ln (1+z)$ is the redshift evolution of the observed galaxy population with the galaxy number density $\bar n_g$, $p=-0.4 \rmd\log\bar n_g/\rmd\log L$ is the slope of the luminosity function, $\delta z_v$ and $\delta z_\chi$ are the lapse in the observed redshift in the matter comoving and the longitudinal gauges, $V$ is the line-of-sight velocity, $\delta r_\chi$ and $\mathcal{K}$ are radial and angular displacements, and $\delta\mathcal{D}_L$ is the fluctuation in the luminosity distance.
Hereby, the subscripts $v$ and $\chi$ denote the dark matter comoving ($v=0$) and conformal Newtonian (zero-shear $\chi=0$) gauge, respectively.

In addition to the modulation of the galaxy number density due to the matter fluctuation $b\,\dm$, the physical origin of the numerous terms in Eq.~(\ref{eq:obsgalfluct}) are the volume and source distortions.
Since we define the observed galaxy samples in terms of the observed redshift and flux, the fluctuations in the observed redshift and the flux give rise to the source effects $e\,\delta z_v$ and $5p\,\delta\mathcal{D}_L$.
The metric potential and velocity terms define the Lorentz frame of the source galaxies.
The remaining terms in Eq.~(\ref{eq:obsgalfluct}) arise due to the volume distortion, as the volume element we assign in observations based on the observed redshift and angle is different from the volume occupied by the observed galaxies. The volume distortion is decomposed as the radial $\delta r_\chi$ and the angular~$\mathcal{K}$ displacements and the distortion in the observed redshift. 
The notation is written in a physically straightforward manner.

The application of the relativistic description to the modified gravity theories and dark energy models is straightforward.
Although the full relativistic formula in Eq.~(\ref{eq:obsgalfluct}) is obtained by solving the photon geodesic equation in GR, the sole assumption that leads to Eq.~(\ref{eq:obsgalfluct}) is that the space-time is described by the FLRW metric, and photons follow the geodesic. This assumption is valid for all modified gravity and dark energy models considered in this paper.
Therefore, Eq.~(\ref{eq:obsgalfluct}) is applicable to those models without modification.
We further simplify Eq.~(\ref{eq:obsgalfluct}) for the power spectrum analysis by ignoring the complication of survey geometry and projected contributions to $\Delta_{\rm g}^{\rm obs}$.
The former can be readily dealt with once the survey boundaries are specified, while the latter is negligible in the power spectrum analysis~\cite{yoo:10,hui:07}.
In our metric notation, the observed galaxy fluctuation becomes
\bqa
 \Delta_{\rm g}^{\rm obs} & = & b \, \dm - \mu_k^2 k_H \vm  - e \frac{\vm}{k_H} - \Phi'-(5p-2)~\Phi \nonumber \\
 & & + \left[ e + 1 + \frac{H'}{H} + (5p-2) \left( 1 - \frac{1}{a \, H \, r} \right) \right] \Psi \nonumber \\
 & & + \left[ e+\frac{H'}{H} + (5p-2) \left( 1 - \frac{1}{a \, H \, r} \right) \right] i \, \mu_k \vm, \nonumber \\
 & & \label{eq:dgrel}
\eqa
where $\mu_k$ is the cosine of the angle between the Fourier mode and the line-of-sight direction.
Note that we do not equate the two metric potentials $\Psi$ and $-\Phi$, as they differ in modified gravity scenarios ($g\neq0$).

The galaxy fluctuation in Eq.~(\ref{eq:dgrel}) is written in terms of metric perturbations in the longitudinal gauge.
Note that since the full relativistic formula in Eq.~(\ref{eq:obsgalfluct}) is gauge-invariant, it can be evaluated at any choice of gauge conditions, including the longitudinal gauge in Eq.~(\ref{eq:dgrel}).
The relation of the metric potentials to the matter density fluctuation $\dm$ and its velocity $\vm$ depends on the particular gravitational model.

We shall first consider the case of a $\Lambda$CDM universe.
Using the Einstein equations and the conservation equations in GR, the galaxy fluctuation in Eq.~(\ref{eq:dgrel}) can be scaled with the matter density fluctuation as~\cite{yoo:10,jeong:11,yoo:12} 
\bq
 \Delta^{\rm obs}_{\rm g}=\Delta_{\rm g}^{\rm Newt.} + \frac{\mathcal{P}}{k_H^2} \dm - i \, \mu_k \frac{\mathcal{R}}{k_H} \dm, \label{eq:reldg}
\eq
where the two additional terms,
\bqa
 \mathcal{P} & = & -\frac{3}{2}\Om(a)\left[e+f+{H'\over H} + (5p-2)\left(2-{1\over aHr}\right)\right] \nonumber \\
 & & + e \, f, \label{eq:PGR} \\
 \mathcal{R} & = & \left[e+{H'\over H}+(5p-2)\left(1-{1\over aHr}\right)\right] f, \label{eq:RGR}
\eqa
represent the gravitational potential and the velocity contributions to $\Delta_{\rm g}^{\rm obs}$.
Here, the scale dependence of these contributions is explicitly removed, and we have assumed a $\Lambda$CDM universe.
Moreover, we have defined $\Om(a)\equiv H_0^2\Om\,a^{-3}/H^2$.
The galaxy clustering in redshift space is often described by the Newtonian contribution~\cite{kaiser:87}
\bq
 \Delta_{\rm g}^{\rm Newt.} \equiv b \, \dm - \mu_k^2 k_H \vm = b \, \dm + f \, \mu_k^2 \dm, \label{eq:newt}
\eq
where the logarithmic growth rate of structure is
\bq
 f \equiv \frac{\rmd \ln \dm}{\rmd \ln a} = \frac{\dm'}{\dm} \simeq -k_H \frac{\vm}{\dm}. \label{eq:logf}
\eq
The last relation is derived from the conservation equation in Eq.~(\ref{eq:emcons}) and becomes exact in a $\Lambda$CDM universe with GR, where $\zeta'=0$.
It is shown~\cite{yoo:12} that with the hindsight of the full relativistic formula, the velocity terms in Eq.~(\ref{eq:obsgalfluct}) can be derived by using Newtonian dynamics, but we refer Eq.~(\ref{eq:newt}) to the Newtonian contribution because it is the only non-negligible contribution in the Newtonian limit $k_H\rightarrow\infty$.

In a $\Lambda$CDM universe with GR, the key ingredients $f$, $\mathcal{P}$, and $\mathcal{R}$ in Eq.~(\ref{eq:reldg}) are all scale-independent, simply because the scale dependence of the relation among $\Psi$, $\Phi$, $\vm$, $\dm$ is explicitly taken into account and removed.
However, these coefficients in Eq.~(\ref{eq:reldg}) become scale-dependent in modified gravity and dark energy models.
To properly account for these modifications, we first define the \emph{Newtonian growth rate} as the ratio of the velocity to the matter density,
\bq
 \mathcal{F} \equiv -k_H \frac{\vm}{\dm}, \label{eq:newtf}
\eq
and the conservation equation in Eq.~(\ref{eq:emcons}) yields the logarithmic growth rate of structure
\bq
 f = -3 \frac{\zeta'}{\dm} + \mathcal{F}.
\eq
In the Newtonian limit, where the effective fluid is well-approximated as a single matter fluid with vanishing pressure, the matter comoving-gauge curvature is conserved, $\zeta'=0$, and both quantities are identical, $f=\mathcal{F}$, providing a further justification of its terminology.
However, it is really the ratio of the velocity to the matter density, and the redshift-space distortion in Eq.~(\ref{eq:newt}) constrains~$\mathcal{F}$, not~$f$.
The logarithmic growth rate $f$ can be measured, for example, in the cluster abundance analysis.
However, for the models we consider in~\textsection\ref{sec:DEmodels}, the comoving-gauge curvature can vary in time on large scales, and, hence, the two growth rates $f$ and $\mathcal{F}$ are different.
The deviation due to the comoving curvature is described by the effective fluid as~\cite{PPF:07}
\bq
 \zeta' = -\frac{\kappa^2}{2H^2}~a^2 (\rho_{\rm eff} + p_{\rm eff}) \frac{V_{\rm eff}-\vm}{k_H}.
\eq

By using Eqs.~(\ref{eq:poissoneq}), (\ref{eq:Sigmapoisson}), and~(\ref{eq:newtf}), we derive the relativistic description of galaxy clustering in a PPF universe based on Eq.~(\ref{eq:reldg}),
\bqa
 \mathcal{F} & = & \frac{2k_H^2 f + 9\Om(a) \left[ g' + (g+1)(\Sigma'/\Sigma + f) - 2g \right]  \Sigma}{2k_H^2 - 6 H'/H}, \nonumber\\
 & & \label{eq:FPPF} \\
 \mathcal{P} & = & e \, \mathcal{F} - \frac{3}{2} \Om(a) \Bigg\{g'-2~g+(1+g)\left({\Sigma'\over\Sigma}+f\right) \nonumber \label{eq:PPPF} \\
 & & + (5p-2)(1+g)+(1-g){\mathcal{R}\over\mathcal{F}}\Bigg\} \Sigma, \\
 \mathcal{R} & = & \left[ e + \frac{H'}{H} + (5p-2) \left( 1 - \frac{1}{aH \, r} \right) \right] \mathcal{F}. \label{eq:RPPF}
\eqa

In Figs.~\ref{fig:FST} and \ref{fig:Fphen}, we show the predictions for $\mathcal{F}$ for the different dark energy and gravitational models.
Fig.~\ref{fig:Fphen} illustrates the corresponding model predictions for the relativistic corrections, i.e., the velocity and potential coupling, $\mathcal{R}$ and $\mathcal{P}$, respectively.
We adopt the model and cosmological parameter values used in~\textsection\ref{sec:DEmodels} and the perturbations are obtained employing the linear PPF framework discussed in~\textsection\ref{sec:ppf}.
Note that at fixed redshift, $\mathcal{R}\propto\mathcal{F}$, where for $e=3$ and $p=0.4$, we have
\bq
 \mathcal{R} = \frac{3}{2} \left[ 2 - \Om(a) - (1+w_{\rm eff}+w'_{\rm eff} \ln a) \Omega_{\rm eff}(a) \right] \mathcal{F} \label{eq:RFprop}
\eq
with $\Omega_{\rm eff}(a) \equiv H^{-2} \kappa^2 \rho_{\rm eff}/3$.
Given this proportionality, simply being the consequence of $\mathcal{R}$ describing the coupling to the velocity $k_H V_{\rm m}$ in Eq.~(\ref{eq:dgrel}) and $\mathcal{F}\Delta_{\rm m}=-k_H V_{\rm m}$, we can directly relate the signatures and corresponding measurement significances for the modified gravity and dark energy cosmologies in $\mathcal{F}$ and $\mathcal{R}$.


\subsection{Measurement significance}\label{sec:significance}

\begin{table*}
\begin{tabular}{lcc|cccc|ccc|ccc}
\hline
\hline
\multicolumn{3}{c|}{Models}
& \multicolumn{4}{c|}{PPF function $g(a,k)$}  & 
\multicolumn{3}{c|}{PPF function $\Gamma(a,k)$}  & 
\multicolumn{3}{c}{Signatures} \\
\hline
Name & Type & $w_{\rm eff}$ & $\gqs$ & $\gsh$ & $c_g$ & $n_g$ & $f_G$ & $f_{\zeta}$ & $\cG$ & $\left|\frac{\Delta\mathcal{F}}{\sigma_{\mathcal{F}}}\right|$ & $\left|\frac{\Delta\mathcal{R}}{\sigma_{\mathcal{R}}}\right|$ & $\left|\frac{\Delta\mathcal{P}}{\sigma_{\mathcal{P}}}\right|$ \\
\hline
$\Lambda$CDM     & GR$+\Lambda$ & $-1$           & 0                               & 0              & $\cdot$ & $\cdot$ & $\cdot$ & $\cdot$ & $\cdot$ & $\cdot$ & $\cdot$ & $\cdot$ \\
FDSB             & GR$+\scal$   & $\frac{H^2(\scal')^2 -2U}{H^2(\scal')^2+2U}$ & 0                               & 0              & $\cdot$ & $\cdot$ & 0 & cal. & 0.36 & 0.9 & 2.8 & 0.1 \\
SUGRA            & GR$+\scal$   & $\frac{H^2(\scal')^2 -2U}{H^2(\scal')^2+2U}$ & 0                               & 0              & $\cdot$ & $\cdot$ & 0 & cal. & 0.36 & 0.3 & 0.7 & 0.0 \\
DOOM             & GR$+\scal$   & $\frac{H^2(\scal')^2 -2U}{H^2(\scal')^2+2U}$ & 0                               & 0              & $\cdot$ & $\cdot$ & 0 & cal. & 0.36 & 0.2 & 1.5 & 0.0 \\
PNGB             & GR$+\scal$   & $\frac{H^2(\scal')^2 -2U}{H^2(\scal')^2+2U}$ & 0                               & 0              & $\cdot$ & $\cdot$ & 0 & cal. & 0.28 & 0.1 & 1.3 & 0.0 \\
CPL-1            & GR$+\scal$   & $w_0$          & 0                               & 0              & $\cdot$ & $\cdot$ & 0 & cal. & 0.33 & 0.2 & 1.0 & 0.0 \\
CPL-2            & GR$+\scal$   & $-a$           & 0                               & 0              & $\cdot$ & $\cdot$ & 0 & cal. & 0.36 & 0.3 & 0.1 & 0.0 \\
\hline
$f(R)$           & MG           & $-1$           & $-\frac{1}{3}$                  & cal.            & $\sqrt{1+f_R}\frac{H}{M}$ & 2 & $f_R$     & cal. & 1    & 4.2 & 2.1 & 0.2 \\
NCBD             & MG           & $\frac{\frac{2}{3}\frac{H'}{H}+1}{\Om(a)-1}$ & $\frac{\scal-1}{(2\omega_0-2)\scal}$ & cal.       & $\sqrt{\scal}\frac{H}{M}$ & 2 & $\scal-1$ & cal. & 0.65 & 2.7 & 1.1 & 0.1 \\
sDGP             & MG           & $\frac{\frac{2}{3}\frac{H'}{H}+1}{\Om(a)-1}$ & $\frac{-1}{3-2Hr_c\left(3+\frac{H'}{H}\right)}$ & $\frac{9+\frac{4.59}{Hr_c-1.08}}{8Hr_c-1}$ & 0.14 & 3 & 0 & $0.4\gsh$ & 1 & 3.2 & 2.3 & 0.1 \\
nDGP             & MG           & $\frac{\frac{2}{3}\frac{H'}{H}+1}{\Om(a)-1}$ & $\frac{-1}{3+2Hr_c\left(3+\frac{H'}{H}\right)}$ & $-\frac{1}{2Hr_c+1}$ & 0.4  & 3 & 0 & $0.4\gsh$ & 0.15 & 1.3 & 1.0 & 0.0 \\
PHEN-1           & MG           & $-1$           & 0                               & $g_0a^3$       & 1    & 2 & 0 & 0 & 1 & 0.1 & 0.0 & 0.0 \\
PHEN-2           & MG           & $w_0+(1-a)w_a$ & $g_0$                           & $g_0$          & $\cdot$ & $\cdot$ & 0 & 0 & 1 & 1.7 & 1.6 & 0.1 \\
\hline
\hline
\end{tabular}
\caption{Example models of gravitational and quintessence theories studied in~\textsection\ref{sec:DEmodels} with their corresponding PPF functions and parameter describing the linear perturbations (see~\textsection\ref{sec:ppf}).
Hereby, `cal.' refers to the application of our calibration technique (\textsection\ref{sec:STPPF}).
In the last column, we present {\it approximate} significance levels of the deviations of the modified gravity and dark energy models based on the constraints that may be obtained in a multitracer analysis of future galaxy survey data at $z=0\sim1$ (see~\textsection\ref{sec:significance}).
For simplicity assuming fixed cosmological parameters and scale independence in $\mathcal{F}$, $\mathcal{R}$, and $\mathcal{P}$, the measurement significance of the relativistic correction $\mathcal{R}$ is comparable to its counterpart for $\mathcal{F}$, which in $\Lambda$CDM corresponds to the logarithmic matter-density growth rate $f$.
Thus, relativistic corrections to the Newtonian description of galaxy clustering need to be taken into account for deriving consistent model constraints in a multitracer analysis of future galaxy-redshift survey data.
This conclusion holds even in the case of including more small-scale modes in the analysis, which increases the measurement significance of deviations in $\mathcal{F}$ by a factor of $\sim6$.
Note that for $f(R)$ gravity, we have used $|f_{R0}|=10^{-1}$. Other model parameters are given in~\textsection\ref{sec:DEmodels}.
}
\label{tab:models}
\end{table*}

In a multitracer analysis~\cite{seljak:08} of galaxy-redshift survey data, when the observed galaxies are divided into different samples of different values of galaxy bias $b_i$, the observation of modes transverse to the line-of-sight direction ($\mu_k=0$) yields a measurement of the relation between the different galaxy biases, i.e., $b_i$ up to the normalization $\bar{b}$, as is evident in Eq.~(\ref{eq:newt}). Neglecting stochasticity between the dark matter and the galaxy fluctuation fields, subsequent observations of a mode at $\mu_k\neq0$ give a measurement of $f/\bar{b}$, which is free of sample variance. One can, furthermore, determine the bias, free of cosmic variance, by cross-correlating galaxy-redshift survey data with weak gravitational shear fields~\cite{pen:04}.
A combination of these two methods, therefore, in principle, yields a measurement of the logarithmic growth rate $f$, which is free of sample variance~\cite{mcdonald:08}.

In practice, however, galaxies are discrete objects, and their measurements necessarily involve intrinsic shot noise, not to mention other errors associated with the measurements.
Once the sample variance limit is eliminated, the shot-noise errors are the dominant uncertainties.
Since the dark matter halos in which the galaxies reside have finite size in their extent, their sampling errors deviate from the Poisson noise of point sources. Based on this observation, proper mass-dependent weights can be obtained to approximate dark matter halos as the dark matter distribution, reducing the stochasticity between them.
This shot-noise canceling technique~\cite{seljak:09,hamaus:10,cai:10,hamaus:12} can be combined with the multitracer method to maximize the advantage of both methods.

Bernstein and Cai~\cite{bernstein:11, cai:11} performed a Fisher-matrix analysis  to derive the cosmological constraining power from a large-scale galaxy survey combined with weak lensing measurements.
Adopting the Kaiser formula in Eq.~(\ref{eq:newt}) and applying the multitracer method with the shot-noise canceling technique, they find that in a future half-sky galaxy-redshift survey, including halos of mass $M>10^{10}h^{-1}~M_{\odot}$ and modes with $k<0.03~h/{\rm Mpc}$ at $z=0.5\pm0.1$, assuming that weak lensing can determine the bias to an uncertainty of $\sigma_b=0.01$, the growth index $\gamma$ can be measured to an uncertainty of about $9\%$~\cite{bernstein:11}, where $f\simeq\Om(a)^{\gamma}$.
Note that if including smaller scales, i.e., $k<0.1~h/{\rm Mpc}$, assuming that the linear Kaiser formula still holds and perfect knowledge of the bias, the measurement significance becomes $\sigma_{\gamma}/\gamma\simeq0.015$.
We extrapolate these results to a full-sky galaxy-redshift survey, assuming a scale-independent $\mathcal{F}=f$, to estimate the measurement significance of $\mathcal{F}$ to about $\sigma_{\mathcal{F}}/\mathcal{F}\simeq0.05$ for $k<0.03~h/{\rm Mpc}$, which can be a factor of 6 times stronger for $k<0.1~h/{\rm Mpc}$.

A similar multitracer analysis with the shot-noise cancellation technique but without weak lensing measurements has been performed~\cite{yoo:12} to estimate the measurement significance of the general relativistic coefficients $\mathcal{R}$ and $\mathcal{P}$ in galaxy clustering.
On large scales, the proper relativistic formula in Eq.~(\ref{eq:obsgalfluct}) must be used in galaxy clustering and in the power spectrum analysis, Eq.~(\ref{eq:obsgalfluct}) is conveniently expressed as Eqs.~(\ref{eq:dgrel}) and~(\ref{eq:reldg}) with the two coefficients in Eqs.~(\ref{eq:PGR}) and~(\ref{eq:RGR}).
In an all-sky galaxy survey at $z=0\sim1$ with $k<0.03~h/{\rm Mpc}$, it is found~\cite{yoo:12} that the galaxy samples with $e=3$ and $p=0.4$ can constrain the general relativistic effects to
\bq
 \sigma_{\mathcal{R}}/\mathcal{R}\simeq0.1 \ \ \ \ \textrm{and} \ \ \ \ \sigma_{\mathcal{P}}/\mathcal{P}\simeq10.0
\eq
if all the halos of $M>10^{10}h^{-1}~M_{\odot}$ can be used.

Compared to the constraint on $\sigma_\gamma/\gamma$ in the former analysis, the constraints on the relativistic effects in the latter analysis are weaker, simply because only modes close to the horizon scales are sensitive to the relativistic effects and also because the former analysis makes use of additional information from weak lensing measurements.
Note that the measurement significance in $\mathcal{R}$ is about half of what is expected for $\mathcal{F}$.
However, since the two quantities are proportional with an enhancement of $\mathcal{R}$ over $\mathcal{F}$ given by Eq.~(\ref{eq:RFprop}) that is different for each model and roughly of order $3/2\sim3$, the deviation in $\mathcal{R}$ may be enhanced, and the measurement significance thereof, $|\Delta\mathcal{R}/\sigma_{\mathcal{R}}|$, can become comparable to $|\Delta\mathcal{F}/\sigma_{\mathcal{F}}|$ or even exceed it (see Figs.~\ref{fig:FST}-\ref{fig:RPphen} and Table~\ref{tab:models}).

To \emph{approximately} estimate the measurement significance of deviations in the relativistic galaxy-density fluctuation, Eq.~(\ref{eq:reldg}), in nonstandard cosmologies from its $\Lambda$CDM counterpart, for simplicity, we assume \emph{scale independence} in $\Delta\mathcal{F}$, $\Delta\mathcal{R}$, and $\Delta\mathcal{P}$, i.e., we adopt the subhorizon limit of the corresponding model at all scales.
More specifically, we quote the measurement significance using predictions at $k\simeq20H_0$ and $z=0$.
Note that this is a good approximation for the quintessence models, for which these coefficients are only very weakly dependent on scale, while they are completely scale-independent in the case of $\Lambda$CDM.
We summarize our results in Table~\ref{tab:models}, in which we fix the cosmological and model parameters to the values defined in~\textsection\ref{sec:DEmodels}, assuming that they are accurately determined, for instance, by {\it Planck}.
In this simplified approach, the deviation of up to $>20\%$ from the $\Lambda$CDM prediction that we observe in the velocity coupling term $\mathcal{R}$ in certain quintessence and modified gravity models (see Fig.~\ref{fig:RPphen}) may be detectable at the $>2\sigma$-level, which is comparable to the measurement significance expected in the deviations $\Delta\mathcal{F}$ (see Table~\ref{tab:models}).
Thus, the relativistic corrections contain additional information on gravity and dark energy, which needs to be taken into account in consistent horizon-scale tests of departures from $\Lambda$CDM using horizon-scale galaxy clustering.
This conclusion even holds in the case of including more small-scale modes ($k<0.1~h/{\rm Mpc}$) and perfect knowledge of bias for constraining $\mathcal{F}$.

However, we stress again that in our estimation of the detection significance, we restrict to a simplified description of the modified gravity theories and dark energy models, neglecting the strong scale dependence of $\mathcal{F}$, $\mathcal{R}$, and $\mathcal{P}$ at $k\lesssim10H_0$, which we observe in a number of modified gravity cosmologies.
A proper relativistic description within the PPF framework is, therefore, essential for constraining those models on horizon scales.
We leave a more sophisticated analysis of the implications of these scale dependencies on the constraints inferred from a multitracer analysis of future galaxy-redshift surveys to future work.
In this context, it will be of great interest to evaluate whether modifications of the nature described by, e.g., the PHEN-1 model, which has been designed to reproduce $\Lambda$CDM at quasistatic subhorizon scales but shows deviations at near-horizon scales, will be distinguishable from the concordance model through a horizon-scale test of the kind described here.


\section{Conclusions} \label{sec:conclusion}

We explore the signatures of modified gravity and dark energy models in the Newtonian~\cite{kaiser:87} and relativistic contributions~\cite{yoo:09,yoo:10,bonvin:11,challinor:11,bruni:11,baldauf:11,jeong:11,bertacca:12} to galaxy clustering.
Thereby, we adopt the Hu-Sawicki PPF formalism~\cite{PPF:07}, which unifies the linear perturbations around the FLRW background for a large class of modified gravity and dark energy models and provides a consistent and efficient framework to determine the linear perturbations from the near-horizon to the quasistatic subhorizon scales.
As example cosmologies, we consider a number of quintessence models, $f(R)$ gravity, a Brans-Dicke universe with dynamical Brans-Dicke parameter and scalar-field potential, both branches of the DGP model, and phenomenological modifications of gravity.
Since for general scalar-tensor theories, a PPF description in the framework of~\cite{PPF:07} has not been formulated previously, we develop a calibration procedure for finding PPF functions that determine the linear fluctuations of minimally and nonminimally coupled scalar-tensor gravity models.
We test our approach with a number of quintessence and extended quintessence models, finding good agreement between the predictions of the PPF perturbations and the model-specific fluctuations.
In the limit of $f(R)$ gravity, which is equivalent to a scalar-tensor model with Brans-Dicke parameter $\omega=0$, we find that our fitting prescription reproduces the known $f(R)$ PPF fits of~\cite{PPF:07}.
Our PPF calibration fits for general scalar-tensor theories provide an efficient procedure for consistently computing the associated perturbations on all linear scales.
They improve upon the widely-used simple quasistatic subhorizon approximation that can be applied to obtain perturbations of scalar-tensor models at large scales, which, however, breaks down near the horizon and cannot correctly describe the clustering of the dark energy associated with the scalar field of the minimally coupled models on large scales and its additional horizon-scale anisotropic stress in the nonminimally coupled cases.

We have applied the PPF formalism to galaxy clustering, providing a proper relativistic treatment on horizon scales, where the relativistic effects become substantial.
As in GR, the relativistic formula of galaxy clustering in modified gravity theories can be characterized by three parameters: the Newtonian growth rate~$\mathcal{F}$, the velocity contribution~$\mathcal{R}$, and the potential contribution~$\mathcal{P}$.
In modified gravity, the Newtonian growth rate $\mathcal{F}=-k_H\vm/\dm$ slightly differs from the usual logarithmic growth rate of structure $f$ and the redshift-space distortion in galaxy clustering measures $\mathcal{F}$, not the usual~$f$.
Compared to GR, in modified gravity and dark energy models, the three quantities $\mathcal{F}$, $\mathcal{R}$, and $\mathcal{P}$ have different values and exhibit scale dependencies.

We approximately estimate the detection significance of these effects in a multitracer analysis of future all-sky galaxy-redshift survey data employing halos of mass $M>10^{10}h^{-1}~M_{\odot}$ at low redshifts.
Such an analysis may have the potential of discriminating between the relativistic correction sourced by the velocity coupling produced in a number of quintessence or modified gravity scenarios and their counterpart from the concordance model.
Thereby, we find a significance level that is comparable to the measurement significance expected in the deviations of the logarithmic matter-density growth rate.
Thus, we conclude that the relativistic corrections contain additional information on gravity and dark energy, which needs to be taken into account in consistent horizon-scale tests of departures from the concordance model using horizon-scale galaxy clustering.

Note that we restrict to a simplified estimation of the measurement significance, neglecting strong scale dependencies in the velocity fluctuation at the largest measurable scales, which appear in a number of modified gravity cosmologies.
In this context, we further design modifications of the concordance model, which reproduce the standard relations at quasistatic subhorizon scales and cannot be tested by conventional cosmological probes.
The multitracer analysis, however, may provide a method to discriminate between those scenarios.
We leave a more sophisticated analysis of the parameter constraints that may be obtained from future galaxy surveys to future work.
Such an analysis will include effects from additional scale dependencies in the growth and in the relativistic coefficients as well as an evaluation of the employment of these effects to discriminate between horizon-scale deviations and the concordance model.


\section*{Acknowledgments}

We thank Daniele Bertacca, Nico Hamaus, Roy Maartens, Francesco Pace, and Daniel Thomas for useful discussions. LL and KK are supported by the European
Research Council.
JY is supported by the SNF Ambizione Grant.
KK is also supported by the STFC (grant no. ST/H002774/1) and the Leverhulme trust.
Numerical computations have been performed with Wolfram $Mathematica^{\rm \tiny \textregistered}~8$.


\appendix

\section{Scalar-tensor gravity} \label{app:sttheory} \label{app:stbackground} \label{sec:STpert}

For completeness, in the following, we shall review the most important aspects of minimally and nonminimally coupled scalar-tensor gravity of the kind defined by~Eq.~(\ref{eq:action}) to our study.
We begin by discussing the cosmological background equations.
In order to calibrate the linear PPF perturbations for scalar-tensor theories and for the comparison of the performance of their corresponding fits, we also need to determine the full model-specific fluctuations.
For this purpose, we briefly review the linear perturbation theory of the scalar-tensor gravity models.
We refer to~\cite{espositofarese:00} for a more comprehensive discussion of this modified cosmological perturbation theory.
For simplicity, we restrict the background and linear perturbations to a spatially flat matter-only universe.


\subsection{Minimally coupled models} \label{sec:QSpert}

We describe minimally coupled scalar-tensor models by the modified Einstein-Hilbert action Eq.~(\ref{eq:action}) with $F(\scal)=Z(\scal)\equiv1$.
Variation of the action with respect to the metric $g_{\mu\nu}$ and the scalar field $\scal$ yields the Einstein field equation
\bqa
 R_{\mu\nu} - \frac{1}{2} g_{\mu\nu} R & = & \kappa^2 T_{\mu\nu} - g_{\mu\nu} U \nonumber \\
  & & + \partial_{\mu} \scal \partial_{\nu} \scal - \frac{1}{2} g_{\mu\nu} \partial_{\alpha} \scal \partial^{\alpha} \scal \label{eq:QSEinstein}
\eqa
and the conservation equation
\bq
 \Box \scal = \rmd U/\rmd \scal.
\eq
Furthermore, energy-momentum conservation implies $\nabla_{\mu} T^{\mu}_{\;\nu} = 0$, where $T^{\mu\nu} \equiv 2 \delta\Sm/\delta g_{\mu\nu}/\sqrt{-g}$.

Minimally coupled scalar-tensor models do not modify gravity since the Einstein equations for the Jordan frame metric are not modified, i.e., the Einstein tensor relates to the energy-momentum tensor as in GR and the scalar field can be interpreted as an effective fluid contribution to the matter and radiation components.
The background field equation in a spatially flat FLRW universe are
\bqa
 3 H^2 & = & \kappa^2 \rho + \frac{H^2}{2} \scal'^2 + U, \label{eq:quintFriedmann1} \\
 -2 H \, H' & = & \kappa^2 (\rho + p) + H^2 \scal'^2.
\eqa
Furthermore, the energy-momentum conservation of the scalar field and matter gives
\bqa
 H^2 \scal'' + (H\,H' + 3 H^2) \scal' & = & - \frac{\rmd U}{\rmd \scal}, \\
 \rho' & = & - 3 (\rho + p). \label{eq:enmomconsQS}
\eqa
The energy density and pressure associated with the scalar field also satisfy Eq.~(\ref{eq:enmomconsQS}).
By using the Friedmann equation and the energy-momentum conservation, we obtain the equation of state
\bq
 w_{\scal} \equiv \frac{p_{\scal}}{\rho_{\scal}} = \frac{(H\,\scal')^2 - 2 U}{(H\,\scal')^2 + 2 U}, \label{eq:quintwscal2}
\eq
which remains in the regime $w_{\scal} \in [-1,1]$ for $U>0$.


After some algebraic manipulation, the perturbed Einstein equations, scalar-field equation, and energy-momentum conservation equations around the spatially flat FLRW background become
\bqa
 \Phi' & = & -\Phi - \frac{\kappa^2}{2H^2} \rho \frac{\vm}{k_H} - \frac{\scal'}{2} \dscal, \\
 \dm' & = & \left[ -k_H^2 + \frac{3}{2} \frac{\kappa^2}{H^2} \rho + 3 \frac{H'}{H} \right] \frac{\vm}{k_H} + \frac{3}{2} \scal' \dscal, \\
 \vm' & = & -k_H \Phi - \vm, \\
 \dscal' & = & \left( \frac{2 k_H^2}{\scal'} - \scal' \right) \Phi - \frac{\kappa^2 \rho}{H^2 \scal'} \dm - \left[ \frac{U'}{H^2 \scal'^2} + 3 \right] \dscal. \label{eq:mincouppoisson} \nonumber \\
 & & \label{eq:mincouppoisson}
\eqa
Given the background, $H$ or $w_{\rm eff}$, $U$, and $\scal'$, the evolution of these perturbation variables are solved by numerical integration over $\ln a$ with the initial conditions at early time $a_{\rm i} \lesssim 0.01$, corresponding to the matter-dominated era in GR,
\bqa
 \Delta_{\rm m, i} & = & \frac{2}{3} k_H^2 \Phi_{\rm i}, \\
 V_{\rm m, i} & = & -\frac{2}{3} k_H \Phi_{\rm i}, \\
 \delta\scal_{\rm i} & = & 0, \\
 \Phi_{\rm i} & = & \frac{3}{5} \zeta_{\rm i},
\eqa
where the comoving-gauge curvature $\zeta_{\rm i}$ is a constant.


\subsection{Nonminimally coupled models} \label{sec:BDpert}

We describe nonminimally coupled scalar-tensor models in the modified Einstein-Hilbert action, Eq.~(\ref{eq:action}), in their Brans-Dicke representation, Eq.~(\ref{eq:bdparam}).
Variation of the action with respect to the metric $g_{\mu\nu}$ and the scalar field $\scal$ yields the modified Einstein and scalar-field equations,
\bqa
 \scal \left( R_{\mu\nu} - \frac{1}{2} g_{\mu\nu} R \right) & = & \kappa^2 T_{\mu\nu} + \nabla_{\mu} \partial_{\nu} \scal - g_{\mu\nu} \Box \scal - g_{\mu\nu} U \nonumber \\
 & & + \frac{\omega}{\scal} \left( \partial_{\mu} \scal \partial_{\nu} \scal - \frac{1}{2} g_{\mu\nu} \partial_{\alpha} \scal \partial^{\alpha} \scal \right), \nonumber \\
 & & \label{eq:BDEinstein}
\eqa
and
\bq
 2 \frac{\omega}{\scal} \Box \scal = - R - \frac{\rmd}{\rmd \scal} \left( \frac{\omega}{\scal} \right) \partial_{\alpha} \scal \partial^{\alpha} \scal + 2 \frac{d U}{d \scal}. \label{eq:BDscalarfield}
\eq
We further require energy-momentum conservation, $\nabla_{\mu} T^{\mu}_{\;\nu} = 0$.

Note that nonminimally coupled scalar-tensor models modify gravity since the Einstein equations for the Jordan frame metric are not in their general relativistic form, i.e., the Einstein tensor does not relate to the energy-momentum tensor as in GR.
The background modified Einstein equations and scalar-field equation are
\bqa
 3 \scal \, H^2 & = & \kappa^2 \rho + \frac{H^2}{2} \frac{\omega}{\scal} \scal'^2 - 3 H^2 \scal' + U, \label{eq:BDfriedmann1} \\
 -2 \scal \, H \, H' & = & \kappa^2 (\rho + p) + H^2 \frac{\omega}{\scal} \scal'^2 + H^2 \scal'' \nonumber\\
 & & + (H\,H' - H^2) \scal',
\eqa
and
\bqa
 3 \left( H\,H' + 2H^2 \right) & = & \frac{\omega}{\scal} \left[ H^2 \scal'' + (H\,H' + 3 H^2) \scal' \right] \nonumber\\
 & & - \frac{d}{d \scal} \left( \frac{\omega}{\scal} \right) \frac{H^2}{2} \scal'^2 - \frac{\rmd U}{\rmd \scal}.
\eqa
In addition, we have the usual energy-momentum conservation of matter and radiation
\bq
 \rho'= - 3 (\rho + p). \label{eq:BDenmomcons}
\eq


The perturbed parts of the modified Einstein equations are (cf.~\cite{espositofarese:00})
\bqa
 \Phi + \Psi & = & - \frac{\dscal}{\scal}, \\
 -2 \scal (\Phi' - \Psi) + \scal' \Psi  & = & \frac{\kappa^2 \rho}{H^2} \frac{\vm}{k_H} + \frac{\omega}{\scal} \scal' \dscal - \dscal + \dscal', \nonumber\\
\eqa
as well as
\bqa
  - 3 \scal' \Psi' & = & (2 k_H^2 \scal - \frac{\omega}{\scal} \scal'^2 + 3 \scal') \Psi + \frac{\kappa^2 \rho}{H^2} \dm \nonumber\\
 & & + \left[ \frac{1}{H^2} \frac{d U}{d\scal} + \left( k_H^2 - 6 - 3\frac{\scal'^2}{\scal^2} \right) + 3 \frac{\omega}{\scal} \scal' \right. \nonumber\\
 & & + \left. \frac{1}{2} \frac{d}{d\scal} \left( \frac{\omega}{\scal} \right) \scal'^2 \right] \dscal + (\omega + 3)\frac{\scal'}{\scal} \dscal'. \nonumber\\
 & &
\eqa
From the perturbed scalar-field equation, one obtains the dilaton fluctuation equation
\begin{widetext}
\bqa
 \dscal'' + \left( \frac{H'}{H} + 3 + \frac{\omega'\scal - \scal'}{\omega \, \scal} \right) \dscal' + \left[ k_H^2 + \left( \frac{\omega' \scal - \scal'}{\omega \, \scal \, \scal'} \right)' \frac{\scal'}{2} - 3 \left( \frac{H'}{H} + 2 \right) \frac{1}{\scal'} \left( \frac{\scal}{\omega} \right)' + \frac{1}{H^2 \scal'} \left( \frac{U'}{\omega}\frac{\scal}{\scal'} \right)' \right] \dscal & = & \nonumber \\
 \frac{\scal}{\omega} \left\{ k_H^2 (\Psi+2\Phi) + 3 \left[ \Phi'' + \left(\frac{H'}{H}+4\right) \Phi' - \Psi' \right] \right\} + \scal'(\Psi'-3\Phi') - \frac{2}{H^2}\frac{U'}{\omega}\frac{\scal}{\scal'} \Psi. & &
\eqa
\end{widetext}
Finally, the conservation equations imply
\bqa
 \dm' & = & + 3\left[ - \Phi' + \left( \frac{H'}{H} + 1 \right) \frac{\vm}{k_H} + \frac{\vm'}{k_H} \right] \nonumber\\
 & & -k_H \vm, \label{eq:STcons1} \\
 \vm' + \vm & = & k_H\Psi. \label{eq:STcons2}
\eqa
To solve for the nonminimally coupled scalar-tensor model perturbations, given the background, $H$ or $w_{\rm eff}$, $\scal$, and $U$, we combine the perturbed modified Einstein and energy-momentum conservation equations to obtain a second-order differential equation for $\Phi$, which includes dependencies on $\dscal$.
We combine this with the dilaton equation depending on $\Phi$ and $\dscal$.
We use the perturbations for general $F$, $U$, and $Z$, without rewriting them in the Brans-Dicke representation (see, e.g.,~\cite{espositofarese:00}).
We follow~\cite{sanchez:10} in this combination, however, solving the system of coupled differential equations with respect to $\ln a$ instead of cosmic time.
This produces two coupled second-order differential equations for $\Phi$ and $\dscal$, which we solve with
the initial conditions at early times $a_{\rm i} \lesssim 0.01$ corresponding to the GR limit $\Phi'_{\rm i}=\delta\scal'_{\rm i}=\delta\scal'_{\rm i}=0$, where the initial metric perturbation $\Phi_{\rm i}$ is a constant.

Note that since we use divisions by $Z$, we cannot set $Z=0$ to strictly recover $f(R)$ gravity.
For $f(R)$ gravity, we, therefore, solve the perturbations given in~\cite{song:06} for the designer model with a code developed and tested in~\cite{lombriser:10}.
Furthermore, we use a code, where we solve the nonminimally coupled scalar-tensor perturbations given here in Brans-Dicke representation setting $\omega=0$.
We test the three nonminimally coupled scalar-tensor perturbation theory solvers in the limit in which $\omega\rightarrow0$ and find good agreement between the codes.


\subsubsection*{Numerical computation of the NCBD background} \label{sec:NCBDbackground}

Finally, note that in order to obtain the background for the NCBD model, we follow~\cite{sanchez:10}, assuming matter dominance and using the first Friedmann equation, the scalar-field equation with $\rmd H/\rmd t$ replaced by the second Friedmann equation, and the energy conservation equation, i.e., Eqs.~(\ref{eq:BDfriedmann1}) through (\ref{eq:BDenmomcons}), to solve for $\scal$, $a$, and $\rho_{\rm m}$ as functions of cosmic time $t$.
This produces coupled differential equations that are of second order in $\scal$ and first order in $a$ as well as $\rho_{\rm m}$, which we integrate over $t$ using the initial conditions
\bqa
 t_{\rm i}^2 & = & \frac{4}{9\Omega_{\rm m0}}a_{\rm i}^3, \\
 \kappa^2 \rho_{\rm m, i} & = & \frac{4}{3t_{\rm i}^2},
\eqa
with constant initial scale factor $a_{\rm i}$ and constant $\scal_{\rm i}$ as well as $\dot{\scal}_{\rm i}$ for the scalar field, where the dot represents a cosmic-time derivative.


\section{PPF fits} \label{app:PPFfits}

In the following, we review the PPF fits for $f(R)$ gravity (\textsection\ref{sec:fRgravity}) and the DGP braneworld model (\textsection\ref{sec:DGP}).
Finally, we give details on the numerical integration of the PPF perturbations.


\subsection{PPF for $f(R)$ gravity} \label{sec:fRPPF}

The PPF expressions for $f(R)$ gravity were developed in~\cite{PPF:07} and have been shown to provide excellent fits to the full perturbation evolutions while at the same time making the integration computationally less expensive~\cite{PPF:07,lombriser:10}.
On superhorizon scales, i.e., in the limit of $k_H\rightarrow0$, the metric ratio $g_{\rm SH}=\phip/\phim$ is defined through the metric perturbations from solving
\bqa
 \Phi'' + \left( 1 - \frac{H''}{H'} + \frac{B'}{1-B} + B \frac{H'}{H} \right) \Phi' & & \nonumber \\
 + \left( \frac{H'}{H} - \frac{H''}{H'} + \frac{B'}{1-B} \right) \Phi & = & 0, \label{eq:fRSH}
\eqa
and using the relation
\bq
 \Psi = \frac{-\Phi - B \, \Phi'}{1-B}, \label{eq:fRSHrel}
\eq
where $B$ is the Compton wavelength parameter defined in Eq.~(\ref{eq:compton}).
Eq.~(\ref{eq:fRSHrel}) follows from conservation of curvature fluctuation ($\zeta'=0$) and momentum, considering the superhorizon anisotropy relation $\phip=-BH'a\vm/k$.

In the quasistatic regime, $g_{\rm QS}=-1/3$ and at intermediate scales, $g(a,k)$ can be determined through the interpolation formula Eq.~(\ref{eq:gakPPF}), where $c_g=0.71\sqrt{B}$ and $n_g=2$.
Further, $f_{\zeta}=-g_{\rm SH}/3$, $f_G=f_R$ and $c_{\Gamma}=1$.
Note that the approximation breaks down at intermediate scales and low redshifts for values of $B_0 \gtrsim 1$~\cite{lombriser:10}, which are, however, strongly ruled out by observations~\cite{schmidt:09c,lombriser:10,lombriser:11b}.


\subsection{PPF for DGP} \label{sec:ppf_dgp}

The PPF functions for sDGP were derived in~\cite{PPF:07} (also see~\cite{koyama:05}) and have been shown to provide excellent fits to the full perturbation evolutions while at the same time making the integration computationally less expensive~\cite{PPF:07,fang:08a,lombriser:09}.
They encapsulate bulk effects in an effective $3+1$ description. In~\cite{lombriser:09}, these results were extrapolated to cases with nonvanishing brane tension and curvature.
The PPF functions for nDGP were described in~\cite{seahra:10} based on bulk calculations of~\cite{song:07b} and~\cite{cardoso:07} with brane tension but no curvature.
Simplified fits with additional extrapolation for spatial curvature are given in~\cite{lombriser:09}.
Here, we adopt the fits of~\cite{lombriser:09} assuming spatial flatness $K=0$.

In the quasistatic high-$k$ limit, the DGP model predicts
\bq
 g_{\rm QS} = -\frac{1}{3} \left[ 1 - 2\sigma \, H \, r_c \left( 1 + \frac{H'}{3H} \right) \right]^{-1}.
\eq
On superhorizon scales, the sDGP and nDGP fits are
\bqa
 g_{\rm SH}^{\rm sDGP}(a) & = & \frac{9}{8 H \, r_{\rm c} - 1} \left( 1 + \frac{0.51}{H \, r_{\rm c} - 1.08} \right) \\
 g_{\rm SH}^{\rm nDGP}(a) & = & -\frac{1}{2 H \, r_{\rm c} + 1},
\eqa
respectively.
For the interpolation on intermediate scales through Eqs.~(\ref{eq:Gamma}) and (\ref{eq:gakPPF}), we have $c_g=0.14$ and $c_{\Gamma}=1$ in sDGP and $c_g=0.4$ and $c_{\Gamma}=0.15$ in nDGP.
Furthermore, $n_g=3$, and the PPF functions that relate the metric to the density can be set as $f_{\zeta} = 0.4 g_{\rm SH}$ and $f_G=0$.

Note that since the PPF fits to the full 5D DGP perturbations have been studied in~\cite{PPF:07,seahra:10} and we do not repeat these calculations here, we refrain from showing any DGP predictions for $g(a,k)$ and $\Sigma(a,k)$ and refer to~\cite{PPF:07,seahra:10} for comparisons of the PPF with the full DGP perturbations.


\subsection{Numerical computation of PPF perturbations} \label{sec:ppfcode}


Given the background expansion history $H(z)$ or $w_{\rm eff}(z)$ and the PPF functions and parameter, $g$, $f_G$, $f_{\zeta}$, and $\cG$, we evolve the linear perturbations in the PPF formalism described in~\textsection\ref{sec:ppf} through the conservation equations, Eqs.~(\ref{eq:PPFenergyconservation}) and (\ref{eq:PPFmomentumconservation}), the differential equation for $\Gamma$, Eq.~(\ref{eq:Gamma}), with source Eq.~(\ref{eq:Gammasource}), and the modified Poisson equation, Eq.~(\ref{eq:poissoneq}).
They compose a system of coupled first-order differential equations in $\dm$, $\vm$, and $\Gamma$ as well as an additional constraint equation for $\phim$.
The initial conditions are set at early times $a_{\rm i} \lesssim 0.01$ with
\bq
 V_{\rm m, i} = \left\{ k_H \frac{H}{H'} \left[ (g'+1) \phim' + (1-g+g') \phim \right] \right\}_{a=a_{\rm i}},
\eq
where the initial metric perturbation $\Phi_{-,{\rm i}}$ is a constant and $\Phi_{-,{\rm i}}'=\Gamma=0$.


\vfill
\bibliographystyle{arxiv_physrev}
\bibliography{relppf}

\end{document}